\documentclass[twocolumn,showpacs,preprintnumbers,amsmath,amssymb,pre]{revtex4}
\usepackage{graphicx}% Include figure files
\usepackage{dcolumn}% Align table columns on decimal point
\usepackage{bm}% bold math

\newcommand{\dd}{\mathrm{d}}

\begin{document}

\title{A core genetic module : the Mixed Feedback Loop.}
\author{Paul Fran\c cois and Vincent Hakim}
\affiliation{
Laboratoire de Physique Statistique 
\footnote{ LPS is laboratoire associ\'e aux universit\'es Paris VI and VII.},
CNRS-UMR 8550,
Ecole Normale Sup\'erieure, 24, rue Lhomond 75231 Paris France
},
\date{\today}

\begin{abstract}
 
The so-called Mixed Feedback Loop (MFL) is a small two-gene network 
where  protein A regulates
the transcription of protein B and the two proteins form a
heterodimer. It has been found to be statistically over-represented in statistical analyses of gene and protein interaction databases and to lie at the core
of several computer-generated genetic networks.
Here, we propose and mathematically study a model of the MFL and show that, by
itself, it
can serve both as a bistable switch and as a clock (an oscillator) depending
on kinetic parameters. The MFL phase diagram as well as a detailed description
of the nonlinear oscillation regime are presented and some biological examples
are discussed.
The results emphasize the role
of protein interactions in the function of genetic modules and the usefulness
of modelling
RNA dynamics explicitly.
\end{abstract}

\maketitle

\section{Introduction}

\begin{figure}
\includegraphics[width=0.3\textwidth, keepaspectratio=true]{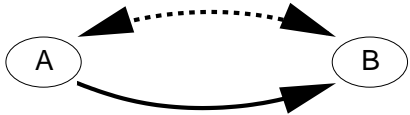}
\caption{\label{motif.fig} Schematic representation of the over-represented
MFL motif \cite{PNASAlon}. The bold arrow represents a
transcriptional regulation interaction and the
dashed double arrow a protein-protein interaction.}
\end{figure}

Biological cells rely on   complex networks of biochemical interactions.
Large scale statistical analyses have revealed that the 
transcriptional regulation networks
and the
networks of protein-protein interaction for different organisms 
are far from random and  contain
 significantly recurring patterns \cite{Milo2002} or motifs. 
Mathematical modelling is useful to determine if these motifs
can by themselves fulfill useful functions and
it has, 
for instance, helped to show that 
 the common 'Feed Forward Loop' motif
in transcriptional regulation \cite{Milo2002} 
can act  as a persistence detector\cite{Mangan2003}.
More recently, a combined analysis of protein-protein interactions
and transcriptional networks in yeast ({\em Saccharomyces cerevisiae}) 
has pointed out several motifs of mixed interactions \cite{Yeger04,PNASAlon}.
The simplest such motif  composed of both transcriptional and protein-protein 
interactions is the two-protein Mixed Feedback Loop (MFL) depicted
in Fig.~\ref{motif.fig}. It is composed of
a transcription factor A, produced from gene $g_a$, and of another protein B , produced from gene
$g_b$.  $A$ regulates the transcription of gene $g_b$ and
also directly interacts with protein $B$. 
The MFL has independently been obtained as the core motif of several networks
produced
in a  computer
evolutionary search  aiming at designing small functional genetic 
modules performing  specific functions
 \cite{Francois2004}. 
Here, to better understand the possible functions
of this basic module,  we propose a model of the MFL 
based on the simplest biochemical interactions. This mathematical model 
is described in section \ref{model.sec}.
This is  used
to show that there exists  wide ranges of kinetic parameters where the
MFL behaves either as a bistable switch or as an oscillator. 
For the convenience of the reader, an overview of the different dynamical
regimes is provided in
section \ref{over.sec}. 
They are delimited in parameter space
and their main characteristics are summarized. 
We then give  detailed descriptions of the bistable regime in section
\ref{switch.sec} and of
the nonlinear oscillations in section \ref{osc.sec}. A comparison is also
made with a simple auto-inhibitory gene model with delay.
 In the concluding section,
the important role played by protein dimerization
and the usefulness of explicitly modelling  mRNA dynamics are underlined and
biological examples of the two proposed functions
of the MFL are discussed.

\section{A mathematical model of the MFL.}
\label{model.sec}
\subsection{Model formulation}
As previously described, the MFL consists of two proteins $A$ and $B$ and
their genes $g_a$ and $g_b$, 
such that $A$ regulates the transcription of gene $g_b$ and also directly
interacts with $B$. Our aim is to analyze the dynamics of this
small genetic module and see what can be achieved in the simplest setting.
Therefore, 
different
cellular compartments and separate concentrations for the
nucleus and cytoplasm are not considered and biochemical reactions are modelled
by simple rate equations.

\begin{figure}
\includegraphics[width=0.4\textwidth, keepaspectratio=true]{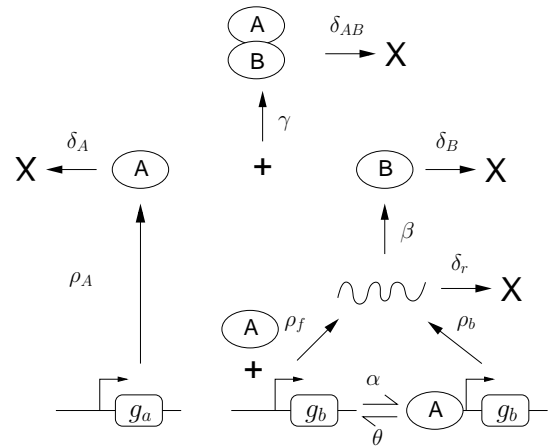}
\caption{\label{model.fig} The proposed model of the
MFL motif. The Greek letters denote the model different kinetic constants.
The large crosses symbolize the degradation of the corresponding species.}
\end{figure}

The proposed MFL model is represented schematically in
Fig.~\ref{model.fig} and consists of  four equations that are described and 
explained
below.
The concentration of a chemical species X is denoted by square brackets
and the
cell volume is taken as volume unit. So,  $[X]$ represents the effective number
of $X$ molecules  present in the cell. 

The first two equations model the transcriptional regulation of
gene $g_b$ by protein A:
\begin{eqnarray}
{\frac{\dd \mathrm{[\it g_b}]}{\dd t} } & {=} & {\theta
  \mathrm{[{\it g_b}:A]}-\alpha\mathrm{[{\it g_b}] [A]} } \label{gene}\\
{\frac{\dd [\mathrm{r_b}]}{\dd t}} & {=} & {\rho_{f}\mathrm{[\it g_b]}+ 
\rho_{b} \mathrm{[{\it g_b}:A]}-\delta_{\mathrm{r}} \mathrm{[r_b]}}  \label{RNA}
\end{eqnarray}
where it is assumed that
gene $g_b$ exists under two forms, 
with A bound to its promoter
with probability $[{\it g_b}:A]$  and without A with probability $[{\it g_b}]$. Since
$\mathrm{[\it g_b}]+[{\it g_b}:A]=1$, the single Eq.~(\ref{gene}) is sufficient
to describe the transition between the two forms. Specifically,
A proteins bind to the $g_b$ promoter at a rate
 $\alpha$ and when bound they are released at a rate $\theta$. 
The regulation of transcription of gene $g_b$
by protein A is described by Eq.~(\ref{RNA}) where $\mathrm{r_b}$ stands for 
$g_b$ transcripts.
When A is bound to $g_b$ promoter,
transcription is initiated at a rate $\rho_{b}$ and otherwise it is initiated
at a rate $\rho_{f}$. Thus, $\rho_{b}>\rho_{f}$ corresponds to transcriptional 
activation by
A and $\rho_{b}<\rho_{f}$ to transcriptional repression. 
A first order degradation for $g_b$ mRNA at a constant 
rate $\delta_{\mathrm{r}}$ has also been assumed. As given,
the description is
strictly valid for a single copy gene. However, it also applies for a gene
with $g_0$ copy (e.g $g_0=2$) provided that
$\rho_f$ and $\rho_b$ are understood to be $g_0$-fold greater than the corresponding elementary rates.

The production of A and B proteins and their complexation are described
by the following two equations that complete our description of the
MFL module, 
\begin{eqnarray}
{\frac{\dd \mathrm{[A]}}{\dd t}} & { =} & { \rho_{\mathrm{A}}-\gamma
\mathrm{[B]\, [A]}-\delta_{\mathrm{A}} \mathrm{[A]}\label{A} } \nonumber\\
&  & { +\theta \mathrm{[{\it g_b}:A]}-\alpha\mathrm{[{\it g_b}][
    A]} } \\
{\frac{\dd \mathrm{[B]}}{\dd t}} & {=} & {\beta\mathrm{[r_b]}
-\gamma \mathrm{[B]}\, \mathrm{[A]} 
-\delta_{\mathrm{B}} \mathrm{[B]} } \label{B}
\end{eqnarray}
where [A] and  [B] respectively denote the  
concentration of  proteins A  and B.
Since, regulation of gene $g_a$  is not considered, separate
descriptions
of its transcription and translation are not needed and it is simply
assumed in Eq.~(\ref{A}) that protein A is produced at a given basal rate
$\rho_{\mathrm{A}}$. 
The second crucial interaction of the MFL, the direct
interaction between protein A and B is taken into account by assuming
that A and B associate at a rate $\gamma$. 
It is  assumed that B  does not interact with a
protein A that is  bound 
to  $g_b$ promoter. In addition Eq.~(\ref{A}) assumes
a first-order degradation of protein A
at a constant rate $\delta_A$, and its last two terms  come
from the (small) contribution to the concentration of A in solution  of the binding (unbinding) of A
 to (from)  $g_b$ promoter.
Eq.~(\ref{B}) assumes that  B proteins 
are produced from the transcripts of gene $g_b$
at a rate $\beta$ and are degraded at a rate $\delta_B$
  in addition to their association with A proteins.

As described by Eq.~(\ref{A},\ref{B}), the
  complexation between A and B proteins 
proceeds at a rate $\gamma$.  
For simplicity, we suppose that the AB complex does not
interact with genes $g_a$ and $g_b$, we  neglect its dissociation
\cite{footdis}
%These assumptions simplify the analysis but do not play a crucial role.
%It was for instance checked that dissociation of the complex with a rate 
%$\gamma_r$ only slightly modify the dynamics of the MFL. Its effect
% can be described
%in the present model 
%to a good accuracy by using an effective  association rate $\gamma$ as briefly
%discussed in section \ref{}.}
% Actually, the quasi static assumptions made 
%in the mathematical analysis remain true and it is possible 
%to see that the dynamics of the dominating protein 
%(and consequently of the whole system) is not changed at the 
%lowest order. However, the introduction of this dissociation 
%modifies the stability of the limit cycle. For instance, 
%with the parameters of Fig. \ref{Oscillation}, 
%taking $\delta_{AB}=\delta_A$, the dissociation rate $\gamma_r$ 
%can be taken up to $0.58 \quad min^{-1}$ without destroying oscillations.} 
and
simply assume that it is fully degraded at a rate
$\delta_{AB}$ after its formation,
\begin{equation}
{\frac{\dd\mathrm{[AB]} }{\dd t}}  {=}  {\gamma \mathrm{[B]\, [A]} -
\delta_{AB}\mathrm{[AB]}}
\label{eqc}
\end{equation}
Since
the complex AB does not feed back on the dynamics of the other species,
its concentration does not need to be  monitored and
Eq.~(\ref{eqc}) is not explicitly considered in the following.

\subsection{Values of kinetic parameters and a small dimensionless parameter}
Even in this simple model, ten kinetic constants should be specified.
It is useful to consider
the possible  range of their values both to
assess the biological relevance of the different dynamical regimes
and to orient   
the model analysis.

Half-lives of mRNA range from a few minutes to several hours and are peaked around
20 minutes in yeast \cite{wang02pnas}. $\delta_r=0.03 min^{-1}$ can therefore be taken
as a typical value.

For the transcription factor-gene promoter interaction, typical values
appear to be
a critical concentration $\theta/\alpha=[A]_0$  
in the nanomolar range, a bound state lifetime of several minutes and 
activated transcription rates of a few mRNAs per minute. 
 We therefore assume $\alpha=\theta/40$,  that
$\theta$ is of the same order as $\delta_r$ and $\rho_f\sim min^{-1}, 
\rho_b\sim min^{-1}$. 

Proteins half-lives vary from a few minutes to several
days \cite{glickman01physiolrev}. The hour appears as a
typical value. We choose $\delta_A=\delta_B=0.01 min^{-1}$ and more
generally consider that the A and B protein half-lives 
are of the same order or longer than that of $g_b$ mRNA.
For protein production, $\beta=3$ protein molecules per mRNA molecule 
per minute appears a
plausible value for an eucaryotic cell \cite{albe4}. This gives $\rho_A\simeq 100 min^{-1}$ when combined with the 
previous values for mRNA production.

We assume that protein interaction is essentially limited by diffusion. 
A diffusion constant $D \simeq 2.5 \mu m^2 s^{-1}$.
\cite{Elowitz99}  gives a time $s^2/D$ of about a minute for diffusion
across a cell of a size $s=10\,\mu m$.
We therefore choose $\gamma \simeq 1 \quad min^{-1}$.

It is convenient to adimension Eqs.~(\ref{gene},\ref{A}) to decrease
as far as possible the number of independent parameters.

We first normalize the $g_b$ mRNA concentration by the concentration
that gives a
 production of B
protein equal to that of A. We thus define the dimensionless concentration
$r=\beta [\mathrm{r_b}]/\rho_{A}$. We normalize the protein 
concentrations
by the equilibrium
concentration resulting from production at a rate $\rho_A$ and dimerization
at a rate $\gamma$ and define
 $A=\sqrt{\gamma/\rho_{\mathrm{A}}}\mathrm{[A]}$,
 $B=\sqrt{\gamma/\rho_{\mathrm{A}}}\mathrm{[B]}$.
We also take as time unit the inverse of $g_b$ mRNA degradation rate 
$1/\delta_r$. With theses substitutions, Eqs.~(\ref{gene}-\ref{B})
read,
\begin{eqnarray}
{\frac{\dd g}{\dd t} } & {=} & {\tilde{\theta}\left[(1-g)-\,g \frac{A}{A_0}
\right]  }\label{gsimple}\\
{\frac{\dd r }{\dd t }} & {=} & {\rho_0\, g+ \rho_1 (1-g)-r } \label{rsimple}\\
\frac{\dd B}{\dd t} & {=} & \frac{1}{\delta}(r-A\,B)-  d_{b}B \label{Bsimple}\\
\frac{\dd A}{\dd t} & {=} & \frac{1}{\delta}(1- A\, B) +\mu\tilde{\theta}
\left[(1-g)-\,g\frac{A}{A_0}\right]
- d_{a} A  \label{Asimple}
\end{eqnarray}

where we  have defined  the following rescaled parameters
$\delta=\delta_{\mathrm{r}}/\sqrt{\rho_{\mathrm{A}}\gamma}$ ,
$\rho_1=\beta\rho_{b}/(\rho_{\mathrm{A}}\delta_r)$, $\rho_0=\beta\rho_{f}/(\rho_{\mathrm{A}}\delta_r)$
$\tilde{\theta}=\theta/ \delta_r $,
$\mu=\sqrt{\gamma/\rho_{\mathrm{A}}} $, $d_a= \delta_{\mathrm{A}}/\delta_r$ and
$d_b=\delta_{\mathrm{B}}/\delta_r$. We have also introduced the dimensionless
critical binding concentration $A_0=\sqrt{\gamma/\rho_A}\,\theta/\alpha$.
The model still depends on seven parameters. In order to simplify its analysis,
it is useful to note that $\delta$ is a small parameter (approximately equal
to $3 \times 10^{-3}$ with the previous estimations). The influence of three 
key parameters of order one 
in Eq.~(\ref{gsimple}-\ref{Asimple}) is particularly examined in the following.
These are 
 $\rho_0$ and
$\rho_1$ which measure the strengths of the two possible states of B production
(with or without A bound to gene $g_b$) as compared to that of A, and
$\tilde{\theta}$ which  compares the rates of A unbinding from DNA to 
that of mRNA degradation ($\alpha$ is supposed to vary with $\theta$ to 
maintain a fixed critical binding concentration $A_0$).

\section{Overview of the dynamics in different parameter regimes.}
\label{over.sec}
We provide here an overview of the different dynamical regimes of
the MFL which are depicted in Fig.~\ref{passage} and summarize their
characteristics.
The behavior of the MFL  depends on whether protein A is a
transcriptional activator or a transcriptional repressor and on the
strengths of the two transcription rates of gene $g_b$ (i.e. with A bound
or not to its promoter). More precisely, the key parameters are
 the strengths of 
B protein production, $\beta \rho_f/\delta_r$ and 
$\beta \rho_b/\delta_r$,
as compared to the production rate $\rho_A$ of protein A. It is therefore
convenient to consider the previously introduced ratios
$\rho_0=\beta  \rho_f/(\delta_r \rho_A)$ and 
$\rho_1=\beta  \rho_b/(\delta_r \rho_A)$.

We have observed that depending on the values of $\rho_0$ and $\rho_1$
the MFL can be monostable, exhibit bistability or display oscillations.
We qualitatively describe these
three different cases in the following. Their biological relevance is further
discussed in section \ref{disc.sec}.
\subsection{Monostable steady states}
The simplest case
occurs when the production rate of A is higher or lower than the production
rate of B, irrespective of the state of $g_b$ promoter. That is when
both $\beta \rho_f/\delta_r$ and $\beta \rho_b/\delta_r$ are either
higher  or lower than $\rho_A$. In this case, the MFL has a single stable state
to which it relaxes starting from any initial conditions. 

When both B production
rates are higher than the production rate of A (i.e. $\rho_0>1$ and $\rho_1>1$),
A proteins
are quickly complexed by B and are unable to interact with $g_b$ promoter.
The concentration $[A]$ of uncomplexed A proteins is therefore 
low and results from a simple 
balance between production and complexation. 
The high concentration of uncomplexed
$B$ proteins is the effective result from transcription at
the free $g_b$ promoter rate, 
complexation and degradation,
\begin{equation}
[B]\simeq \frac{1}{\delta_B}\left(\beta \frac{\rho_f}{\delta_r}-
\rho_A\right), \ \ \ [A]\simeq \frac{\rho_A}{\gamma [B]}
\label{bhighmono}
\end{equation}

An equally simple but opposite result holds when both B production rates
are smaller than the production rate of A ($\rho_0<1$ and $\rho_1<1$). 
Then, the concentration of uncomplexed A is high , $g_b$ promoter is occupied
by A and a low B concentration results from a balance between complexation and
degradation
\begin{equation}
[A]\simeq \frac{1}{\delta_A}\left(\rho_A-\beta  \frac{\rho_b}{\delta_r}
\right), \ \ \ [B]\simeq \frac{\beta   \rho_b}{ \delta_r [A]\gamma }
\label{blowhmono}
\end{equation}

The dynamics of the MFL is richer when the production rate of A is intermediate
between the two possible production rates of B, $\beta \rho_f/\delta_r$ and 
$\beta \rho_b/\delta_r$  i.e. when either
 $\rho_0>1>\rho_1$  or $\rho_1>1>\rho_0$. We consider these two cases in turn. 

\subsection{Transcriptional repression and bistability.}
We first discuss the case when A is a transcriptional repressor,
$\rho_0>1>\rho_1$. Then, two stable steady states can coexist.
Let us suppose first that no A is bound to $g_b$ promoter.
Then the production rate of B is larger than the production rate
of A, and all produced A proteins are quickly complexed. This stably prevents
the binding of A proteins to  $g_b$ promoter  and maintain a steady state
with  low  A and  high B concentrations approximately equal to
\begin{equation}
 [B]_1\simeq \left(\beta \frac{\rho_f}{\delta_r}-\rho_A \right)
\frac{1}{\delta_B},\ \ \ [A]_1\simeq \frac{\rho_A}{\gamma [B]_1} ,
\end{equation}
The second opposite possibility is that A is sufficiently abundant to
repress the transcription of gene $g_b$. Then, since the production
rate of A has been supposed to be higher than the production rate of B
in the repressed state, B proteins are quickly complexed but uncomplexed
A proteins are present to maintain the repression of gene $g_b$ 
transcription. This gives rise to a second stable state 
with high A and low B concentrations
approximately equal to
\begin{equation}
[A]_2\simeq \left(\rho_A -\beta  \frac{\rho_b}{\delta_r}\right)
\frac{1}{\delta_A},\ \ \ [B]_2\simeq \frac{\beta \rho_b}{\gamma [A]_2 
\delta_r}
\end{equation}

The bistability domain is only exactly given by the simple inequalities
$\rho_0>1$ and $\rho_1<1$ when the ratio $\delta$ of
$g_b$ mRNA degradation rate over the effective protein dynamics rate
is vanishingly small. As shown in section \ref{switch.sec} and on 
Fig.~\ref{passage}, for small
value of $\delta$, it is more accurately given by
\begin{eqnarray}
\rho_0 &>& 1 + 2\sqrt{(1-\rho_1)\, \delta_B/A_0},\ \rho_1\le 1 ,
\nonumber\\
\rho_1&<& 1-2\sqrt{(\rho_0-1)\, A_0\, \delta_A},\ \rho_0\ge 1.
\end{eqnarray}
In this intermediate range of production of A proteins, the network reaches
one or the other fixed points, depending on its initial condition. 
The
existence of this bistability domain can serve to convert a graded increase in
A production in a much more abrupt switch-like response in A (and B) 
concentration as
shown in Fig.~\ref{switch.fig}. The usefulness of this general feature of 
multistability has been recently discussed in different contexts 
\cite{ferrell02, ozbudak04nat}.

\subsection{Transcriptional activation and oscillations.}
\label{oscsur}
When A is a transcriptional activator, the complexation
of B with A acts as a negative feedback and can serve to diminish
the variation in B protein concentration when A varies. This leads
also to oscillations when A
production  lies in the
intermediate range $\rho_1>1>\rho_0$.
This oscillatory behavior mainly depends on the ratios of protein production
$\rho_0, \rho_1$ and exists for a large range of DNA-protein
interaction kinetics . However, a faster kinetics leads to a smaller oscillatory domain
as shown in Fig.~\ref{passage}. Oscillations cannot be sustained when
$\tilde{\theta}$ becomes large and comparable to
$ 1/\delta$ (with $A_0$ fixed), 
or equivalently when $\theta\sim\delta_r/\delta$.
It can also be noted that for a sufficiently large activation of
transcription by A, there exists a domain of coexistence between
oscillations and steady behavior
as shown in Fig.~\ref{osczoom.fig}, i.e. depending on initial conditions
concentrations will oscillate in time or reach steady levels.

For small $\delta$, oscillations are nonlinear for most parameters.
As can be seen in Fig.~\ref{Oscillation},
an oscillation cycle comprises two phases in succession, a first phase of 
duration $T_1$ when
the concentration of protein A is high followed by a second
phase of duration $T_2$ 
where the concentration of B is high. 
A full oscillation cycle of  
period $T=T_1+T_2$
proceeds as follows.
Let us start with low concentrations of A and B proteins at the beginning
of the first phase.
When 
no A is bound to $g_b$ promoter, B production rate is
lower than A production rate and complexation cannot prevent the increase
of A concentration. When the concentration of A has reached a critical 
level, A start to bind to $g_b$ promoter and activate transcription. This 
results in a higher  production of B than A and the diminution of free A
by complexation. Since A concentration is high, the produced B's are quickly
complexed and the concentration of uncomplexed B's remain low. 
Eventually, A concentration drops below the binding level
and no longer activates $g_b$ transcription.
This marks the transition between the two parts of the oscillation cycle. B continues for a while
to be produced from the transcripts and since few A's are present
this now leads to a rise of the concentration of free B's. Finally, the 
concentrations of B transcripts and B proteins drop and phase II of the 
oscillation cycle terminates.
A new cycle begins with low concentrations of A and B proteins.

One remarkable feature of the nonlinear oscillations coming from the
smallness of the parameter $\delta$, 
%is that the nonlinear oscillations
%depend in a simple way on several parameters. For instance, 
is that 
the concentrations of $A$ and $B$ proteins in their respective high phase depend
weakly on the complex association rate $\gamma$, as long as
it is not too small for the oscillation to exist, and that the period of the 
oscillations is strikingly insensitive to the exact value of $\gamma$. For
the parameters of Fig.~\ref{Oscillation}, the oscillation period 
only changes by about
$1\%$ when $\gamma$ is reduced from $1 min^{-1}$ to $2\times 10^{-2} min^{-1}$.
This remains true even if the formation 
of the complex $AB$ is not irreversible as supposed in the present model.
We have checked that the  case when the complex $AB$ forms
with an association rate $\gamma_a$,  dissociates with a rate $\gamma_d$
and is degraded with a rate $\delta_{AB}$ is well described 
by the present model when one takes the effective rate
$\gamma=\gamma_a \delta_{AB}/(\gamma_d+\delta_{AB})$ for the irreversible
formation of the complex
(as obtained from a quasi-equilibrium assumption).

In the two following sections, we provide a more detailed analysis of the MFL
different dynamical regimes and derivations of the results  here summarized.

\section{The switch regime.}
\label{switch.sec}
We first determine 
the possible steady states for the MFL. 
The free gene, mRNA and B 
protein concentrations are given in a steady state as a function of A
concentration by,
\begin{eqnarray}
g&=&\frac{A_0}{A+A_0}\label{eqg}\\
r&=&\frac{\rho_1 A+ \rho_0 A_0}{A+A_0}\\
B&=&\frac{\rho_1 A+ \rho_0 A_0}{(A+A_0)(A+\delta d_b)}
\label{eqB}
\end{eqnarray}
The concentration of A itself satisfies 
the following equation 
\begin{equation}
1=\delta d_a A +\frac{(\rho_1 A+ \rho_0 A_0)A}{(A+A_0)(A+\delta d_b)}
\label{eqA}
\end{equation}
For $\delta=0$, the right-hand side (r.~h.~s.~) goes monotonically from $\rho_0$ at small A
to $\rho_1$ at large A and a solution, $A_2=A_0(\rho_0 -1)/(1-\rho_1)$
exists only when $\rho_0<1<\rho_1$ or
$\rho_1<1<\rho_0$. For small $\delta$, two other steady states are possible.
A steady state with a small concentration of A, 
$A_1\simeq \delta d_b/(\rho_0-1)$, exists when $\rho_0>1$. Inversely
a steady state with a large concentration of A, $ A_3 \simeq 
(1-\rho_1)/(\delta d_a)$ exists when $\rho_1<1$. 
Therefore, Eq.~(\ref{eqA}) has multiple (i.e. three) fixed points
only  when A is a transcriptional repressor in the region $\rho_1<1<\rho_0$.
This is the parameter regime which we examine in the rest of this section.

\subsection{The high A state}
We show that the state with a high concentration of A proteins is stable.
The form of
Eqs.~(\ref{Bsimple}) and (\ref{Asimple}) suggests to simplify the
analysis
by using the fact that protein quickly
 reach a quasi-equilibrium state.
However, 
both
proteins cannot be in quasi-equilibrium at the same time. For instance, when
proteins $A\gg B$, only $B$ is in quasi-equilibrium and 
$A$  scale as $1/\delta$.  
When A concentration is high, in the limit of small $\delta$,   we set
$a=\delta A$. Substitution in Eq.~(\ref{gene},\ref{B}) shows 
that  $g$ and $B$ 
reach on a fast time-scale their quasi-equilibrium concentration 

\begin{eqnarray}
{g}&\simeq &
\delta\, A_0/a \\
{B}&\simeq&
\delta\, r/a 
\end{eqnarray}

Therefore the dynamics of the MFL  reduces to the following two
equations in  the limit $\delta\rightarrow 0 $ :

\begin{eqnarray}
{\frac{\dd r }{\dd t }} & {=} & {\rho_1-r} 
\label{ha1}
\\
{\frac{\dd a }{\dd t}} & {=} & {1- r-d_a a} 
\label{ha2}
\end{eqnarray}

Eqs.~(\ref{ha1},\ref{ha2}) clearly show that the high A fixed point is stable
and as found above satisfies $r \simeq \rho_1$, $ B \simeq g\simeq0$,
$a=(1-\rho_1)/d_a$ at the lowest order. This
steady state is possible only if $\rho_1<1$, that is when
B production rate is not high enough to titrate A and to prevent the
repression by A.

\subsection{The high B state.}
The high B state can be analyzed in a very similar way 
in the limit of small $\delta$.
When B concentration is high,  we define
$b=\delta B$. In that case,  $A$ quickly reaches its quasi-equilibrium
concentration:

\begin{eqnarray}
{A}&{\simeq \delta/b} 
\end{eqnarray}

and the MFL dynamics reduces to the following
three equations:

\begin{eqnarray}
{\frac{\dd g}{\dd t} } & {=} & {\tilde{\theta}(1-g) } \\
{\frac{\dd  r }{\dd t }} & {=} & {\rho_0 g+\rho_1 (1-g)-r} \\
{\frac{\dd b }{\dd t}} & {=} & { r-1-d_b b}  
\end{eqnarray}

The concentrations tend toward those of the high B
fixed point  $g\simeq 1$, $r\simeq\rho_0$, $b\simeq(\rho_0-1)/d_b$
$A\simeq 0$. 
This
steady state exists  only if $\rho_0>1$, 
when the production of B proteins is  high enough
to titrate A proteins and to prevent transcriptional repression by A.

\subsection{Bifurcation between the monostable and bistable regimes.
}
The previous analysis has shown that bistability exists in the domain
$\rho_0>1>\rho_1$ when $\delta\ll 1$.
We analyze more precisely the nature and position of the bifurcations between
the monostable and bistable regimes.

\begin{figure}
\includegraphics*[width=0.45\textwidth, keepaspectratio=true]{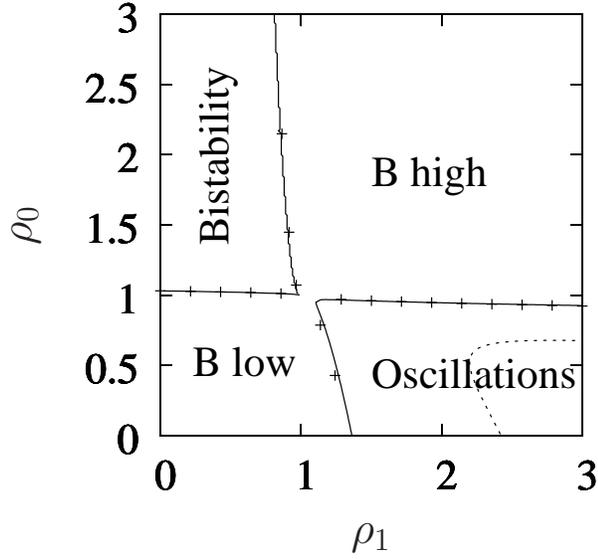}
\caption{\label{passage}
Different dynamical regimes of the MFL as $\rho_0$ and $\rho_1$ are varied.
The borders between different regimes are shown as computed from numerical
solutions of Eq.~(\ref{eqA}) and (\ref{poly}) for
$\tilde{\theta}=1.33$ (full lines) or $\tilde{\theta}=26.6$ (dashed)
as well as given by the asymptotics expansions
[Eqs.~(\ref{bistable1},\ref{bistable2},\ref{osc1}, \ref{osc2})] 
for $\tilde{\theta}=1.33$ ($+$  symbols).
Here, and in all following figures, except when explicitly specified,
the others parameters are
$ [A]_0=40\, mol, \beta= 3 \quad  min^{-1}$,
$\delta_r= 0.03  \quad min^{-1}$, $\delta_{\mathrm{A}}= 0.01 \quad  min^{-1}$,
$\delta_{\mathrm{B}}=
0.01\quad   min^{-1}$, $\rho_A=100 \quad mol.min^{-1}$,  $\gamma= 1 \quad  mol^{-1}.min^{-1}$ where $mol$
stands for molecules and $min$ for minutes.
The corresponding dimensionless parameters are
 $\delta=0.003$, $d_a=d_b=0.33$, $A_0=4$, $\mu=0.31$.
}
\end{figure}

\begin{figure}
\includegraphics*[width=0.4\textwidth,keepaspectratio=true]{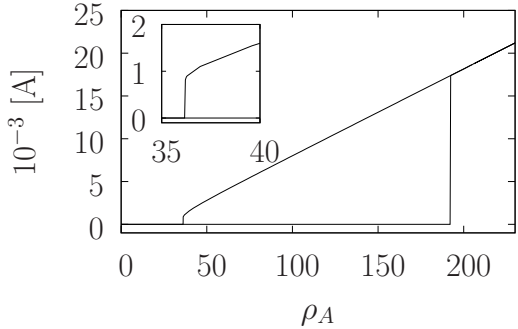}

\caption{\label{Bistable}
In the bistable regime a graded increase of $\rho_A$, the production of A,
results in a jump in A concentration.
The parameters are
$\rho_f=0.2\quad  mol.min^{-1} $ $\rho_{b}=
2 \quad  mol.min^{-1}$, and $\theta=0.04 \quad
min^{-1}$. The other parameters
are as in Fig.~\ref{passage}.
}
\label{switch.fig}
\end{figure}

\begin{figure}
\includegraphics*[width=0.45\textwidth, keepaspectratio=true]{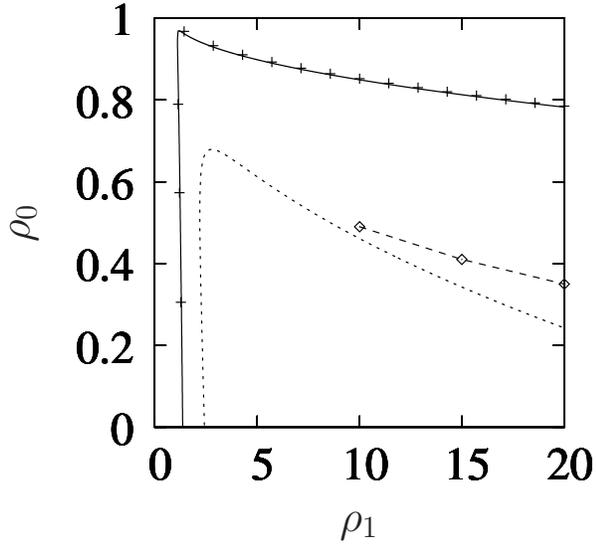}
\caption{\label{osczoom.fig} The oscillation domain for the same parameters
as in Fig.~\ref{passage} but for a larger domain of $\rho_1$, the ratio
of  activated production of B to that of A. As in Fig.~\ref{passage}, the
short-dashed line marks the boundary
of the domain where the steady stable is unstable to oscillations
for $\tilde{\theta}=26.6$.
Numerical simulations (diamonds) show that the oscillating regime is stable
up
 to the long-dashed line. Therefore,
both the limit-cycle and the fixed point are stable
attractors for $\tilde{\theta}=26.6$
in the region between the short-dashed and
long-dashed lines.}
\end{figure}

\begin{figure}
\includegraphics*[width=0.4\textwidth,keepaspectratio=true]{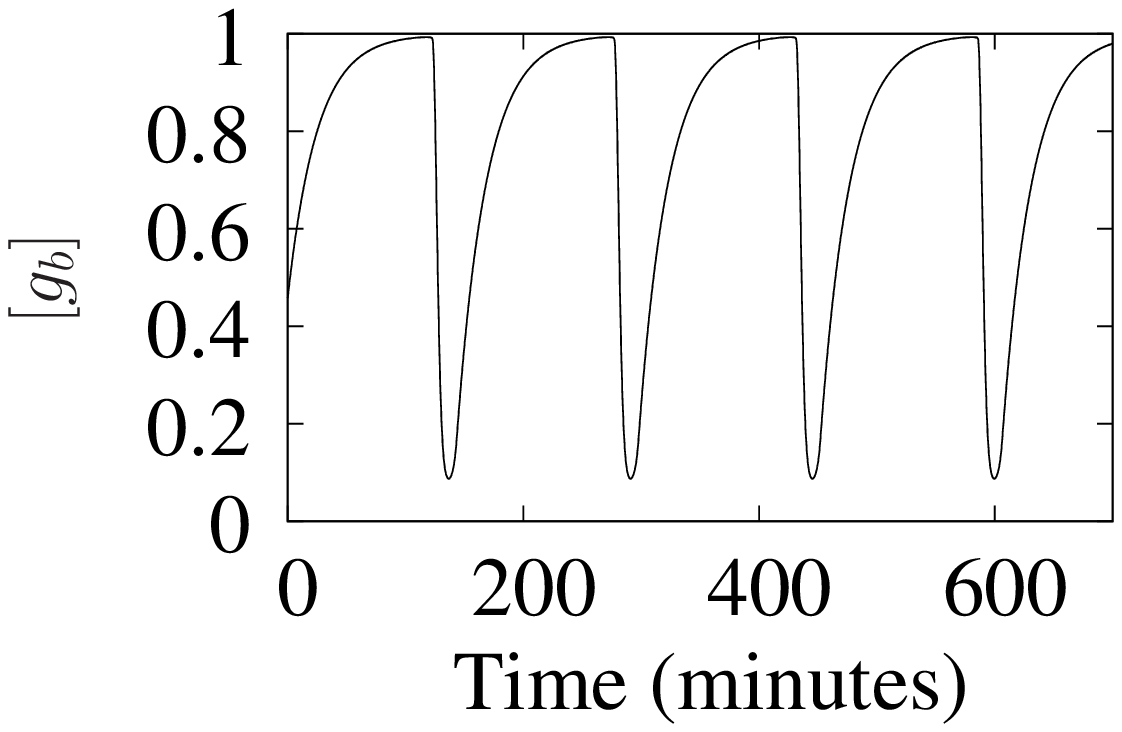}\\
\includegraphics*[width=0.4\textwidth,keepaspectratio=true]{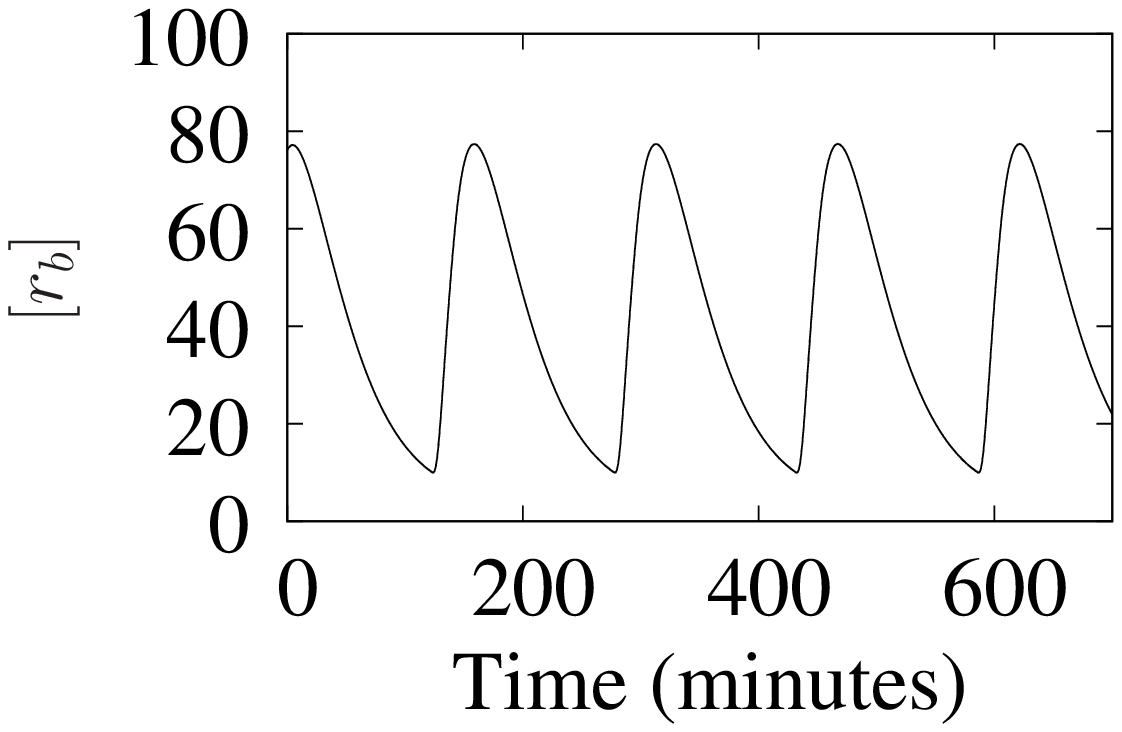}\\
\includegraphics*[width=0.4\textwidth,keepaspectratio=true]{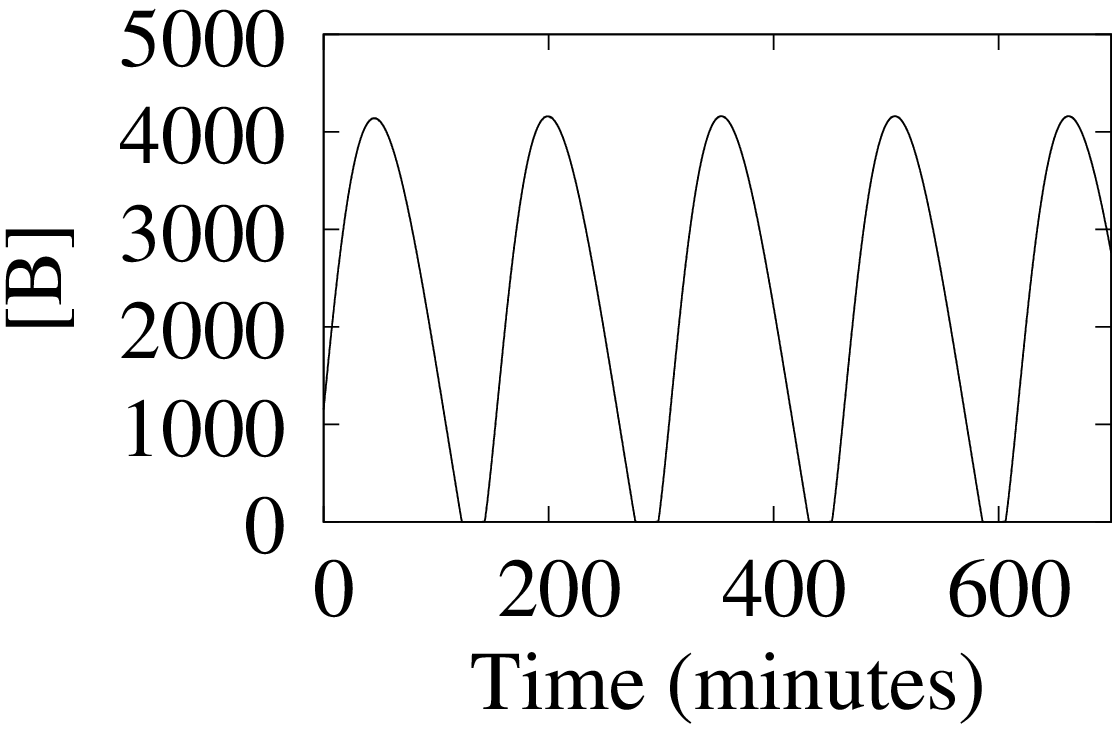}\\
\includegraphics*[width=0.4\textwidth,keepaspectratio=true]{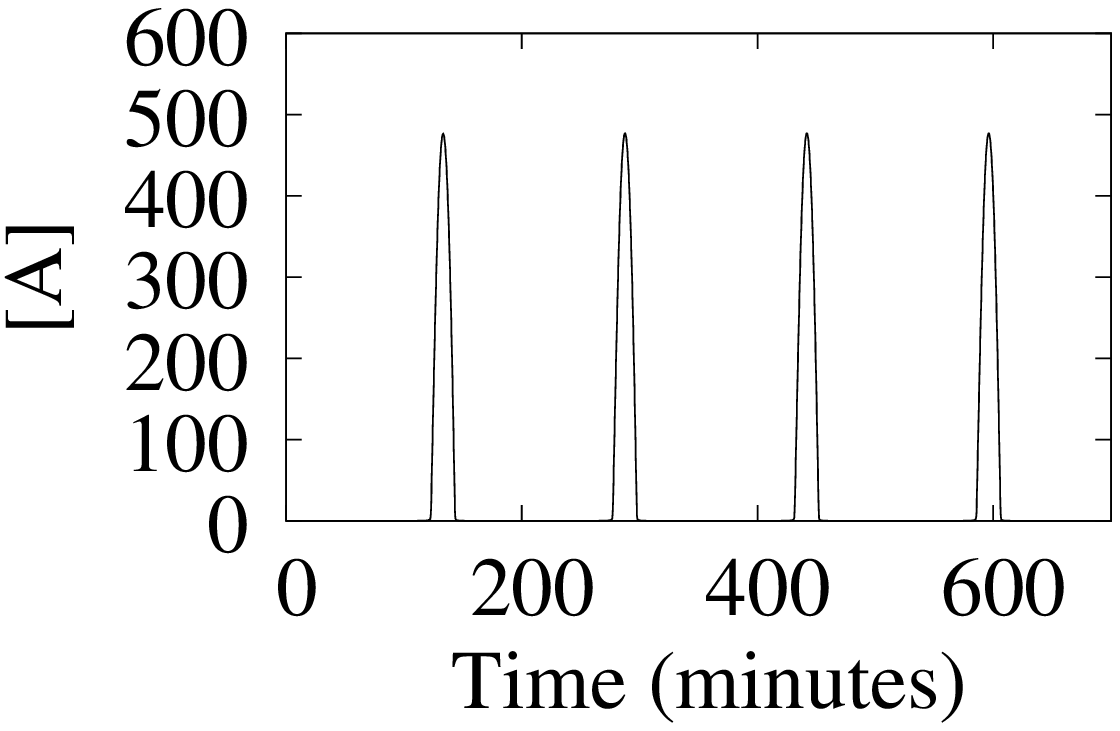}
\caption{\label{Oscillation}  MFL in the oscillatory regime. The concentrations
 of  the different species are shown as  a function of time and display
sustained  oscillations.
Constants are the same as in Fig. \ref{Bistable}
 except that
$\rho_{\mathrm{A}}= 100 \quad mol.min^{-1}$, $\rho_f=0.1\quad
mol.min^{-1} $ $\rho_{b}=5\quad  mol.min^{-1}$.}
\end{figure}

We consider first the transition between the bistable region and
the monostable region with
a high concentration of A (bottom left corner in Fig.~\ref{passage}). It follows from 
Eq.~(\ref{eqA}) that the low ($A_1$) and middle ($A_2$) A concentration
fixed points coalesce and disappear when $\rho_0-1 =O(\sqrt{\delta})$
and both $A_1$ and $A_2$ are of order $\sqrt{\delta}$.
For $A\sim\sqrt{\delta}$
Eq.~(\ref{eqA}) becomes at dominant order after 
expansion, 
\begin{equation}
\rho_0-1=(1-\rho_1) \frac{A}{A_0}+\delta \frac{d_b}{A}
\label{bif1}
\end{equation}
For a given repression strength $\rho_1<1$, the r.~h.~s.~ has a minimum
for $A=\sqrt{\delta d_b A_0/(1-\rho_1)}$ (that is of order $\sqrt{\delta}$
as assumed) such that
\begin{equation}
\rho_0=
1+2 \sqrt{\frac{\delta d_b (1-\rho_1)}{A_0}} +O(\delta) \label{bistable1}
\end{equation}
Above this value of $\rho_0$, Eq.~(\ref{bif1}) has two solutions and
below it has none. Eq.~(\ref{bistable1}) therefore marks the saddle-node
bifurcation line that separates the bistable domain from the monostable domain
with a high A concentration fixed point. For $\delta\rightarrow 0$,
the zeroth order boundary $\rho_0=1$ is of course retrieved.

The approximate expression (\ref{bistable1}) agrees well with an exact
numerical determination of the bifurcation line as shown in
Figure \ref{passage}.

The transition between the bistable regime and the low A concentration
monostable regime (top-right corner of Fig.~\ref{passage}) can be similarly
analyzed. The high ($A_3$)  and middle ($A_2$)
A concentration fixed points can coalesce when for both
the concentration A is of order $1/\sqrt{\delta}$. Expansion of Eq.~(\ref{eqA})
for $A\sim 1/\sqrt{\delta}$ gives
\begin{equation}
1-\rho_1= (\rho_0-1)\frac{A_0}{A} +\delta d_A A
\label{bif2}
\end{equation}
For a given $\rho_0>1$, the r.~h.~s. of Eq.~(\ref{bif2}) is minimum
for $A=\sqrt{A_0(\rho_0-1)/(\delta d_a)}$ and at this minimum $\rho_1$ is equal
to
\begin{equation}
\rho_1=1-2\sqrt{d_a A_0 \delta( \rho_0-1) }\label{bistable2}
\end{equation}
For $\rho_1$ smaller than this value,
 Eq.~(\ref{bif2}) has two roots corresponding to $A_2$ and $A_3$. The two roots
merge and disappear on the saddle-node bifurcation line (\ref{bistable2}) that 
marks the boundary between the bistable domain and the low A concentration
monostable regime.
Comparison between the approximate expression
(\ref{bistable2}) and an exact numerical determination of the bifurcation line
is shown in
Figure \ref{passage}.

\section{The oscillator regime}
\label{osc.sec}
We consider now the case when A is a transcriptional activator, that is
when $\rho_0<\rho_1$. In this case, the MFL has a single steady state
since the right-hand side (r.h.s.) 
of Eq.~(\ref{eqA}) is a monotonically increasing function.
However, as we show below, this steady state is unstable in a large domain of 
parameters and the MFL behaves as an oscillator.
\subsection{The linear oscillatory instability}
We begin by analyzing the linear stability of the fixed point 
$(g^*, r^*, B^*, A^*)$ where $A^*$ is the single solution of Eq.~(\ref{eqA})
and $g^*, r^*$ and $B^*$ are given as functions of $A^*$ by 
Eqs.~(\ref{eqg}-\ref{eqB}). After linearization of the MFL dynamics 
[Eqs.~(\ref{gsimple}-\ref{Asimple}] around this fixed point, 
the complex growth rates $\sigma$ of perturbation growing (or decreasing) exponentially
in time like $\exp(\sigma t)$ are found to be the roots of the following
characteristic polynomial,

\begin{widetext}
\begin{equation}
\left[\sigma+\tilde{\theta}(1+\frac{A^*}{A_0})\right]
\left[\sigma+1\right]
\left[\left(\sigma+d_a+\frac{B^*}{\delta}\right)\left(\sigma+d_b+
\frac{A^*}{\delta}\right)
-\frac{A^*B^*}{\delta^2}
\right]=
\frac{g^*
  \tilde{\theta}}{A_0}\left[\frac{A^*}{\delta^2}
 (\rho_0-\rho_1)- \mu 
  \sigma\left(\sigma+1\right)\left(
\sigma+d_b+\frac{A^*}{\delta}\right)\right] \label{poly}
\end{equation}
\end{widetext}

Again, the fact that $\delta$ is small simplifies the analysis.
We consider first the case when the rescaled concentrations 
$g^*, r^*, B^*, A^*$ are of order one. This corresponds to the fixed
point $A_2$ of the previous part in the parameter regime $\rho_0<1<\rho_1$,
with $A^*\simeq  A_0 (1-\rho_0
)/(\rho_1-1)$, $B^*\simeq 1/A^*$ and  
$g^*\simeq (\rho_1-1)/(\rho_1-\rho_0)$. Let us assume that the roots $\sigma$
diverge as $\delta$ tends to zero. Then, Eq.~(\ref{poly}) reduces at
dominant order to

\begin{equation}
\sigma^3(\delta\sigma+A^*+B^*)=\frac{A^*}{A_0}
\frac{\tilde{\theta}  g^* }{\delta}(\rho_0-\rho_1)
\end{equation} 
Therefore, three roots $\sigma_k, k=0,1,2,$ are of order $\delta^{-1/3}$ and
proportional to the three cubic roots $j^k$ of unity
\begin{equation}
\sigma_k =-j^k \left[\frac{A^*}{A^*+B^*}\frac{g^* }{A_0}
\frac{\tilde{\theta}}{\delta}(\rho_1-\rho_0)\right]^{1/3},\ k=0,1,2.
\end{equation}
The fourth root $\sigma_3$ is of order $1/\delta$ and corresponds 
to a stable mode of real part
$-(A^*+B^*)/\delta$. 
The two roots $\sigma_1$ and $\sigma_2$ are complex conjugate and have
a positive real part. Thus, in this parameter domain, the MFL fixed point
is oscillatory instable. The dynamics tends toward an attractive limit cycle
that we will analyze in the next subsection.
When $A^*$ or $B^*$ grows (i.e when $\rho_1\rightarrow 1$ or $\rho_0\rightarrow 1$) the fixed point instability disappears {\em via} a Hopf bifurcation.
We analyze these two boundaries in turn.

\subsubsection{ $A^*$ high}
When $\rho_1\rightarrow 1$, $A^*$ becomes large and the real parts of
the roots become of order one. Thus, Eq.~(\ref{poly}) needs to be approximated 
in a different way. We neglect $\delta(\sigma+d_b)$ and  
$B^*$ as compared to $A^*$ in Eq.~(\ref{poly}) and obtain
\begin{equation}
(\sigma+\tilde{\theta}\,\frac{\rho_1-\rho_0}{\rho_1-1})(\sigma+1)(\sigma+d_a)=
-\frac{\tilde{\theta}(\rho_1-1)}{\delta
  A_0}
\label{ahosc}
\end{equation}
where we have replaced $A^*$ and $g^*$ by their expressions at the
$A_2$ fixed point.
  When the r.h.s is the dominant constant in Eq.~(\ref{ahosc}),
its three roots are proportional as above to the three roots of (minus) unity
and two of them have positive real parts. On the contrary,
when $\rho_1$ becomes sufficiently close to one, the
r.h.s of Eq.~(\ref{ahosc}) becomes negligible and Eq.~(\ref{ahosc}) has
obviously three real negative roots. Therefore, when $\rho_1\rightarrow 1$, one
traverses the boundary of the linearly unstable region. Two roots of
Eq.~(\ref{ahosc}) traverses the imaginary axis on this boundary in
a Hopf bifurcation.
Assuming (and
checking afterwards) that, on this boundary, these  roots are small 
compared with $1/(\rho_1-1)$, their expressions is found perturbatively
by expanding the first factor in Eq.~(\ref{ahosc}), 
\begin{equation}
\sigma_{\pm}=-\frac{1+d_a}{2}+
\frac{(\rho_1-1)^3}{2(\rho_1-\rho_0)^2 A_0 \tilde{\theta}\delta}
\pm \frac{i}{2} 
\sqrt{\frac{(\rho_1-1)^2}{A_0(\rho_1-\rho_0)\delta}
}
\end{equation}
This provides the location of the stability boundary,
\begin{equation}
\rho_1=1+\left(\delta \tilde{\theta} A_0 (d_a+1)(\rho_1-\rho_0)^2
\right)^{1/3}
\label{osc1}
\end{equation}
In the limit $\delta \rightarrow 0$ the zeroth order
boundary $\rho_1=1$  is of course retrieved. It can be checked that the obtained result
$\rho_1-1\sim \delta^{1/3}$ justifies {\em a posteriori} the use of the $A_2$
$\delta$- independent expression for the fixed point.
This fixed point linear stability boundary (\ref{osc1}) can also be directly
obtained from the Routh-Hurwitz condition \cite{foot}
on the third degree polynomial
(\ref{ahosc}).
Comparison of the small $\delta$ Eq.~(\ref{osc1})
with the exact linear stability boundary is provided in
Fig.~ \ref{passage}.

\subsubsection{ $A^*$ small}
We now consider the upper boundary ($\rho_0\sim 1$) of the linearly
unstable region. 
When $\rho_0$ tends toward one, $B^*$ grows $1/(\rho_0-1)$ and becomes
large compared to
$\delta(\sigma+d_a)$ and $A^*$.
Eq.~(\ref{poly}) then simplifies and reduces at leading order to:

\begin{equation}
(\sigma +\tilde{\theta})
(\sigma+1)(\sigma+d_b)=-E\,
\label{eqla}
\end{equation} 
with 
\begin{equation}
E=
\frac{A^* \tilde{\theta}}{A_0 B^*\delta}(\rho_1-1) 
\label{defe}
\end{equation} 
Again when $E$ is large the roots are proportional to the three roots of
-1 and two have positive real parts. On the contrary,
the three roots are real and
negative when $E$ vanishes. There is a critical value $E_c$ 
such that for $E<E_c$ the fixed point is linearly stable. This occurs
when the real  part of the two complex conjugate roots becomes negative
at the value $E_c$ given by the Routh-Hurwitz criterion \cite{foot}
\begin{equation}
E_c=(d_b+1+\tilde{\theta})(\tilde{\theta}+d_b+d_b\tilde{\theta})-d_b\tilde{\theta}
\end{equation} 
The allied value of $A$ at the fixed point concentration
is obtained from Eq.~(\ref{defe}),
\begin{equation}
A^*=\sqrt{\frac{E_c A_0 \delta}{\tilde{\theta}(\rho_1-1)}} 
\label{eqala}
\end{equation}
Using the fixed point Eq.~(\ref{eqA}), this in turn corresponds to the
line in parameter space
\begin{equation}
\rho_0= 1-\sqrt{\frac{E_c\delta(\rho_1-1)}{ A_0
    \tilde{\theta}}}\left[1-\tilde{\theta}\frac{d_b}{E_c}\right]\label{osc2}
\end{equation}
Eq.~(\ref{osc2}), of course, reduces to the zeroth order boundary
$\rho_0=1$ when $\delta \rightarrow 0$.
Comparison of the small-$\delta$ asymptotic expression
(\ref{osc2})  with the exact linear
stability line is provided in
Fig.~\ref{passage}.

\subsection{A description of the nonlinear oscillations}

The MFL oscillations quickly become strongly nonlinear
away from threshold (or, of course, when the bifurcation is subcritical).
As there is no auto regulatory direct or indirect positive
feedback loops in the MFL, 
two-variable reductions have a negative divergence. It follows
from the well-known Bendixson's criterion \cite{guho}, that they cannot be used
to describe the oscillatory regime. It is for instance not possible to
properly describe the oscillations by only focusing on the protein
dynamics. A  specific
analysis is therefore required and developed in this section.

\begin{figure}
\includegraphics*[width=0.4\textwidth,keepaspectratio=true]{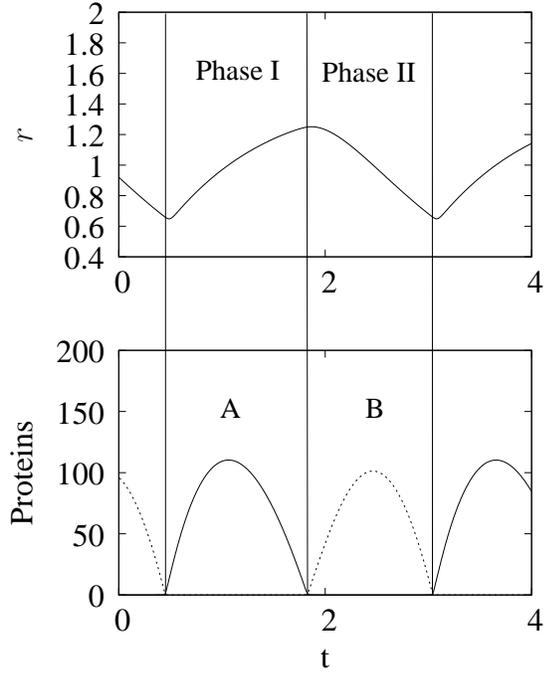}
\caption{\label{phases} Distinction between the two phases for the
  adimensioned equations [Eqs.~(\ref{gsimple}-\ref{Asimple})]. Top panel
  : oscillation of $r$. Bottom panel : oscillations for A (full line)
  and B (dotted). In order to clearly depict phase I and II of an oscillation cycle, the
parameters are here chosen so that the two phases have
  similar durations : $\delta=0.001$, $\rho_1=1.5$,$\rho_0=0$, $\tilde{\theta}=2$,
 $A_0=1$,  $\mu=1$, $d_a=d_b=0$.}
\end{figure}

A full period of the nonlinear oscillations can  be split into two parts: a 
first
phase with $A>>1$,  and a second phase with   $B>>1$ (see Fig. \ref{phases}). 

In the following, these two main phases of the limit
cycle are described. We use simple continuity arguments to match the two phases
and obtain a  description of the whole
limit cycle. The oscillation period is computed (at the lowest-order in $\delta$).  A
full justification of this simple matching procedure and
a detailed study of the transition regimes between the two main
phases are provided in Appendix \ref{appendixA} using matched asymptotics 
techniques.

\subsubsection{Phase I: High A, Low B phase.}
\label{p1ha}
This phase is defined 
as the fraction of the limit-cycle where the concentration of
protein A is larger than $[A]_0$ i.~e. when
$A>>1>>B$.  Eq.~(\ref{Asimple}) leads us to assume that $A$ is of 
the order
$1/\delta$. So we define $a=\delta A$.
In the limit of small $\delta$, as $A$ scales as $1/\delta$, both the 
binding to $g$ promoter and the dynamics of B 
resulting from its complexation with A are fast compared to that of mRNA and
A.
Consequently, $g$ and B are in quasi-equilibrium and obey at lowest order
in $\delta$, 
\begin{eqnarray}
{g}&{=} & \delta {A_0}/{a}\label{gad}\\
{B}&{=} &{\delta r/a }
\end{eqnarray}
The dynamics of the MFL therefore  reduces to the following two
equations at lowest order in $\delta$,
\begin{eqnarray}
{\frac{\dd r }{\dd t }} & {=} & { \rho_1-r} \label{ODErphase1}\\
{\frac{\dd a }{\dd t}} & {=} & {1- r - d_a a} \label{Weffectif}
\end{eqnarray}
The beginning of phase I coincides with the end of phase II where, as
explained below, A concentration is small. Continuity therefore requires
that $a$ vanishes at the start of phase I (a detailed study of the transition
region between the two phases is provided in Appendix \ref{appendixA}).
Denoting by $r_1$, the value of $r$ at the start of phase I, an easy
integration of the linear 
Eq.~(\ref{ODErphase1},\ref{Weffectif}) gives,
\begin{eqnarray}
{r_I(t)} & {=} & { \rho_1+ (r_1-\rho_1) e^{-t}} \label{rphase1} \\
{a_I(t)} & {=} & {\frac{1-\rho_1}{d_a}[1-e^{-d_a t}]
  \!+\!\frac{r_1-\rho_1}{d_a-1}[e^{-d_a t}-e^{-t}]\ } \label{wphase1}
\end{eqnarray}
Subscripts $I$ have been added to $r$ and $a$ in Eq.~(\ref{rphase1}) and
(\ref{wphase1}) to emphasize that the corresponding expressions
are valid during phase I only.
Phase I ends with the fall of A concentration, after a time $t_1$
such that $a_I(t_1)=0$. The mRNA concentration is then equal to 
$r_{I}(t_1)=r_2$.

One can note that the rise and fall of the concentration of protein A 
imposes restrictions on the parameters. The rise  at the beginning of
phase I is possible only if A production dominates over its complexation with
B, that is if $r_1<1$ (Eq.~(\ref{Weffectif}). The following fall requires
the reverse which can only happen if B production becomes sufficiently 
important, namely $r>1$. This requires $r_2>1$ and 
{\em a fortiori} $\rho_1>1$ (Eq.~(\ref{ODErphase1})).

\subsubsection{Phase II : High B phase.}
\label{p2la}
This second 
phase is defined as the part of the limit-cycle where the concentration
of protein A falls below $[A]_0$ i.e.
$B>>1>>A$. The form of Eq.~(\ref{Bsimple}) leads us to assume
that B scales as $1/\delta$ and we define $b=\delta B$.

In the limit of small $\delta$,  the dynamics of $A$ is fast 
compared to that of the other species and A is in quasi-equilibrium.
At lowest order its concentration reads,
\begin{eqnarray}
{A}&{\simeq \delta/b} \label{Welimine}
\end{eqnarray}
Thus, the dynamics of the MFL reduces, at lowest order in 
$\delta$,
to the following
three equations,  
\begin{eqnarray}
{\frac{\dd g}{\dd t} } & {=} & {\tilde{\theta}(1-g) } \label{geffectif}\\
{\frac{\dd  r }{\dd t }} & {=} & { \rho_0 g+\rho_1(1-g)-r} 
\label{rp2}\\
{\frac{\dd b }{\dd t}} & {=} & { r-1-d_b b}  \label{Feffectif}
\end{eqnarray}
Continuity with the previous phase I leads us to require that at the
beginning of phase II $b=0, r=r_2$ and $g=0$ (see Appendix \ref{appendixA}
for a detailed justification).
With these boundary conditions, the linear
Eqs. \ref{geffectif} - \ref{Feffectif} 
can readily be integrated to obtain:

\begin{eqnarray}
g_{II}(t) & {=} & {1-e^{-\tilde{\theta}t}}\nonumber%\label{fcompg}
\\
r_{II}(t) &=& \rho_0+\left[r_2-\rho_0+\frac{\rho_1-\rho_0}{\tilde{\theta}-1}
\right] e^{-t}
-\frac{\rho_1-\rho_0}{\tilde{\theta}-1} e^{-\tilde{\theta} t}\nonumber\\ 
b_{II}(t) &=&  \frac{\rho_0-1}{d_b}[1-e^{-d_b t}]
-\frac{\rho_1-\rho_0}{\tilde{\theta}-1}
\frac{e^{-d_b t}- e^{-\tilde{\theta} t}}{\tilde{\theta}-d_b}\nonumber\\
&+&\left[r_2-\rho_0+
\frac{\rho_1-\rho_0}{\tilde{\theta}-1}
\right] \frac{e^{-d_b t}-e^{-t}}{1-d_b} \label{fcompb}
\end{eqnarray}
Since $b$ vanishes at the beginning of phase II, $b$ should start by rising.
This imposes that B production dominates over complexation
and requires that the concentration $r$ at the beginning of phase II is greater than 1, i.~e.  
$r_2>1$ (see Eq.~(\ref{Feffectif})). Thus, $r$ always remains larger than $\rho_0$ and would 
decrease toward $\rho_0$ for a long enough phase II (Eq.~(\ref{rp2}) and 
(\ref{fcompb})). At the end of phase II,
b should  decrease to
$0$ to continuously match the beginning of phase I. This  requires that
complexation then dominates over production of B  that is $r<1$ (Eq.~(
\ref{Feffectif})),
and it is only possible
when $\rho_0<1$.
With these conditions met,
phase II lasts a time
$t_2$ and ends when $b_{II}(t_2)=0$.

\subsubsection{Matching of the two phases and period determination}
In order to complete the description of the limit cycle, it remains 
to determine the four unknowns $r_1,r_2,t_1$ and $t_2$ from the
four conditions coming from the continuity of proteins and
mRNA concentrations,  
\begin{eqnarray}
r_I(t_1)=r_2,\ a_{I}(t_1)=0,
\label{per1.eq}\\
r_{II}(t_2)=r_1,\ 
b_{II}(t_2)=0. 
\label{per2.eq}
\end{eqnarray}
One possibility to solve these equations is to use Eq.~(\ref{per1.eq})
to express $r_1$ and $r_2$ as a function of $t_1$. It is then
not difficult to see that the third equation 
(\ref{per2.eq}) implicitly determines $t_2$ as a function of $t_1$.
Once $r_1,\,r_2$ and $t_2$ are obtained as functions of $t_1$, the last
equation
$b(t_2)=0$ can be solved for $t_1$  by
a one-dimensional root finding algorithm. In this way, we have determined
$r_1,
r_2, t_1, t_2$ for different sets of kinetic constants and obtained the
rescaled period of oscillation
\begin{equation}
T_r=t_1+t_2 \label{periode},
\end{equation}
the dimensionfull period being  $T=T_r/\delta_r$.
Results 
are shown in Fig. \ref{period}. The analytical period
compares well to results obtained by direct numerical integration of
Eq.~(\ref{gsimple}-\ref{Asimple}) for different values of $\delta$ with, of course,
a closer agreement for smaller $\delta$.

Eqs.~(\ref{per1.eq},\ref{per2.eq}) are difficult to solve analytically but
analytic expressions can be obtained for various limiting cases. For
instance, if the degradation of protein B is negligible ($d_b=0$), 
Eq.~(\ref{fcompb}) simplifies to
\begin{eqnarray}
b_{II}(t) &=&  (\rho_0-1) t
-\frac{\rho_1-\rho_0}{(\tilde{\theta}-1)\tilde{\theta}}
(1-e^{-\tilde{\theta} t})\nonumber\\
&+&\left[r_2-\rho_0+
\frac{\rho_1-\rho_0}{\tilde{\theta}-1}
\right] (1-e^{-t}) \label{b2simp}
\end{eqnarray}
If we further consider the limit of large $\rho_1$, the condition
$b_{II}(t_2)=0$, gives the following
estimate for the duration of phase II, up to exponentially small terms,
\begin{equation}
t_2 \simeq \frac{r_2-\rho_0}{1-\rho_0}+\frac{\rho_1-\rho_0}{\tilde{\theta}(1-\rho_0)},
\end{equation}
as well as the transcript concentration at the end of phase II
\begin{equation}
r_1 \simeq \rho_0\ .
\label{r1simp}
\end{equation}
Similarly, the condition $a_{I}(t_1)=0$ together with Eq.~(\ref{wphase1})
show that $t_1$ is small for large $\rho_1$, 
\begin{equation}
t_1\simeq 2(1-r_1)/\rho_1\simeq 2(1-\rho_0)/\rho_1
\label{t1simp}
\end{equation}
where expression (\ref{r1simp}) has been used in the second equality.
Given the duration (\ref{t1simp}) of phase I, the concentration $r_2$ of
transcript at its end is directly obtained from Eq.~(\ref{rphase1})
and (\ref{r1simp})
\begin{equation}
r_2 \simeq 2-\rho_0 \ .
\end{equation}
This finally provides the estimate of the period for large $\rho_1$ (with
$d_b\simeq 0)$,
\begin{equation}
T_r \simeq t_2 \simeq 2+\frac{\rho_1-\rho_0}{\tilde{\theta}(1-\rho_0)}\ .
\label{Periodeapprox}
\end{equation}

A comparison between Eq.~(\ref{Periodeapprox}) and numerically determined
oscillation periods is provided in Fig.~\ref{period}.

\subsection{Comparison with a simple delayed negative feedback.}

Popular models for genetic oscillators
 consist in a
protein repressing its own production with a phenomenological delay
\cite{jensen03,Lewis2003,monk03}. Using a Hill-function to
model this repression, the simplest model of this kind reads,
\begin{equation}
  \frac{\dd B(t)}{\dd t}=\frac{\rho}{1+[B(t-\tau)/B_0]^n}-c B
\label{delayed}
\end{equation}
where $\rho$ is the protein production rate, $c$ the protein degradation rate
and $\tau$  the phenomenological delay for repression.
The phase diagram of this simple oscillator can be computed 
\cite{Glass88}. In the limit of long delays ($ c\tau\gg 1$), the oscillations are nonlinear. 
In an expansion in the small parameter $\epsilon=1/c\tau$, their period $T$ is 
\begin{equation}
T/\tau=2+\frac{\kappa}{c\tau}+\cdots
\end{equation}
where $\kappa$ is a constant which is approximately equal to 2 for $n=2$ and
$\rho/c$ of order one \cite{Lewfoot}.

It is interesting to note some analogies between this simple model 
with delay and the previously studied MFL. 
When $\tau>>1/c$, the period of the delayed model scales
as $\tau$   while the period of the MFL scales as $\delta_r^{-1}$. The mRNA
half-life in the MFL thus plays the role of the phenomenological delay in
Eq.~(\ref{delayed}). This is in line
with the mRNA  being the major pacemaker of the MFL oscillator . The
rescaled parameter $\epsilon$ in the simple model with delay
  plays a role similar to
the rescaled parameter $\delta$ in the MFL in the sense that, first, for high
values of $\epsilon$, oscillations disappear \cite{Glass88},  and for small
values of this parameter, the period of the oscillations is independent of
$1/c\tau$ at dominant order. In the delayed model, $\epsilon$ represents the
ratio between the typical life-time of protein over the delay, while in the MFL,
$\delta$ represents the ratio between the typical time scale of protein
production and sequestration over the typical life-time of RNA.
These analogies show  that for the MFL $\sqrt{\rho_A \gamma}$ plays the
role of 
the protein degradation constant $c$ 
in the delayed model. Thus protein sequestration by complexation is
the MFL analog  
of protein degradation in the simple model with delay.
The fastness of both these processes
relative to a slow mechanism (delay $\tau$, RNA dynamics) ensures the flipping
of both oscillators between two states (two concentrations of $B$ for
the delayed
model,  A high and B high for the MFL).
\section{Discussion}
\label{disc.sec}

We have proposed a simple model of the Mixed Feedback Loop, an over-represented
motif in different genomes. We have shown that by itself this motif
can serve as a bistable switch or generate oscillations.

\subsection{Importance of dimerization and of RNA dynamics}

The MFL dynamics crucially depends on the post-transcriptional interaction
between A and B proteins. The dynamics of dimerization is fast and allows
a high concentration of only one of the two proteins, thus effectively creating
a dynamical switch between the two species. If, for instance, A proteins are present
at a  higher concentration than B proteins, all B's are quickly titrated and 
there only remains free A and AB dimers.

A bistable system is obtained by coupling this post-transcriptional mechanism
to a transcriptional repression. The created 
bistable switch is quite different from the classical 'toggle switch' 
\cite{toggle} which is based on reciprocal transcriptional repressions between
two genes. In the MFL model, the single transcriptional repression of gene $g_b$ is sufficient
for a working switch.
Moreover, dimerization bypasses the need for cooperativity in 
the toggle switch \cite{Cherry00}, and may render the present module 
simpler to implement in a biological system.

When the dimerization between A and B is coupled to a transcriptional activation, 
the system behaves as an oscillator
if the 
production of A proteins is intermediate between the two possible for B.
Then, 
the level of $g_b$ transcripts controls 
A protein
dimerization  and, of course, B production rates.
At the beginning of the derepression
phase (phase I), the low level of transcripts leads  to a small production
of B proteins and to a rise of A concentration. This first phase ends when
the subsequent high production and accumulation
of  $g_b$ transcripts give rise to a production of B proteins sufficient
for the titration of all free A's.
During the ensuing repression
phase (phase II),  $g_b$ transcription rate is low and the
transcripts degradation produces a continuous decrease of their concentration.
The 
parallel decrease of B production rate ultimately leads 
to the end of this second
phase when B production rate is no longer sufficient for the titration
of all produced free A's.

In summary, the core mechanism of the clock is built by
coupling the rapid switch at the protein level to the slow mRNA dynamics.
The protein switch in turn controls mRNA dynamics {\it via} 
transcriptional activation. 

It is interesting to note that the MFL model in the simple form here analyzed
provides a
genetic network that oscillates without the need for several
specific features previously considered in the literature for related networks.
As already mentioned, oscillations require neither an additional
 positive feedback loop
\cite{TYSONBiophys}, nor self-activation of gene A \cite{vilar02pnas}
nor  a highly cooperative
binding \cite{RuoffRensing} of transcription factors to gene promoters.
Similarly, an explicit delay is not needed although
delays of various origins have been considered
in  models of oscillatory genetic networks in different contexts, 
as further discussed below. 

Besides the motif topology,
one feature of the present MFL model that renders this possible is
the explicit description of transcription and translation.
The condensed representation 
of protein production by a single effective process that is often used, would
necessitate supplementary interactions to make oscillations possible. 
The description of mRNA was for instance omitted in the first version of
a previously proposed evolutionary algorithm
\cite{Francois2004}. Oscillatory networks with a core MFL motifs were 
nonetheless produced but much less easily than in a refined second version
that incorporates mRNA dynamics. This role of mRNA dynamics may well be
worth bearing in mind when considering the elaborate regulation of translation
that is presently being uncovered \cite{richter05nat}.
\subsection{Biological MFL oscillators and switches.}
Given the proposed functions for the MFL motif, it is of primary interest
to know whether cases of its use as an oscillator or a switch are already 
documented. The oscillator function seems the clearest. 
Circadian clocks are genetic oscillators which generate endogenous 
rhythms with a period close to 24 hours and which 
are locked to an exact 24 hours
period by the alternation of day and night.
A MFL motif is found at the core of 
all known eucaryotic circadian oscillators. Taking for illustrative purposes
the fungus {\it 
Neurospora crassa}, one of the best studied model organisms,
it has been established that a dimer of White-Collar proteins (WCC)
plays the role
of A and activates the transcription of the {\it frequency} gene, the 
gene $g_b$ in this example. In turn, the FRQ protein interacts with WCC and prevents
this activation \cite{NeuroDunlap}. Analogous motifs are found in the
circadian genetic networks of flies and mammals, as shown in Table \ref{bioex.tab}.
The p53/Mdm2 module \cite{pAlon} provides a different example. 
The p53
protein is a key tumor suppressor protein and an important role 
is played in its 
regulation by the Mdm2 protein. On the one hand, 
p53, similarly to the A protein, binds to Mdm2 gene
an activates its transcription. On the other hand, the Mdm2 protein, as the B
protein, binds to p53 and both blocks transcriptional activation by p53 and 
promotes its rapid proteolytic degradation. Cells exposed to stress have
indeed been observed to present oscillations in both p53 and Mdm2 levels 
\cite{pAlon}.
In both these examples, many other genetic interactions exist
besides the above described MFL motifs and have been thought to
provide delays necessary for oscillations. 
These have been postulated to come from 
chains of
phosphorylations in circadian networks \cite{GoldPNAS} or from a yet
unknown intermediate species \cite{pAlon} in the p53-Mdm2 system. 
The realization that the MFL
can lead to oscillations without further interactions may be useful
in suggesting alternative models of these particular genetic networks or
in reassessing the role of known interactions. For instance, this has 
led one of us 
to formulate a model of the {\it Neurospora crassa} circadian clock 
\cite{FrancoisBiophys} in which kinases and phosphorylation
influence the cycle period
by modifying proteins degradation rates but are not needed to create key delays.
The prevalence of the MFL motif probably arises from the usefulness of
negative feedback but it may also imply that oscillations in genetic network
are more common than usually thought.
 
The evidence for uses of the MFL motif as a switch appears less clear-cut at
present. The classic case of
the {\it E. Coli} lactose operon \cite{monod61} can be thought as an effective version
of a bistable MFL.  The lac repressor represses the {\it lac} gene and
plays the role of A in the MFL. The {\it lac} gene  directs the
production of a membrane permease which itself drives the  absorption of
external lactose. Since allolactose binding to the lac repressor blocks
its transcriptional activity, allolactose effectively plays the same role as B
in the MFL. The search for a more direct switch example has led us to consider
the different MFL motifs in yeast as reported in \cite{Yeger04} and 
reproduced in Table \ref{bioex.tab}.
It was noted in \cite{Yeger04} that most of these MFL motifs were central
modules in biochemical pathways.
Most of them take part 
in the  cell cycle or in differentiation pathways, which certainly are processes
where the cell switches from one function to another. However, we have found
it difficult to disentangle the MFL's from numerous other known interactions
and to confidently conclude that any of these motifs implements the
proposed switch function. Similar difficulties have been recently emphasized
for general motifs determined on purely statistical grounds and have led
to question their functional significance \cite{mazurie05}. In the present
case, the difficulty of identifying biological cases of MFL switches, 
of course arises from our own very partial 
knowledge of the networks listed in Table I, but it may also be due 
to the fact that
the possible role of the MFL module as a switch 
was not fully realized in previous investigations. 
The present study 
will hopefully help and trigger direct experimental 
investigations of these questions.

\begin{table}
\begin{tabular}{|c|c|c|}
\hline
{ A Species} & { B Species} & {Main biological functions}\\ 
\hline
\multicolumn{3}{|c|}{Bistable systems}\\
\hline
{Lac Repressor} & {Allolactose} &{ Lactose metabolism}\\
\hline
\hline
\multicolumn{3}{|c|}{Oscillators}\\
\hline
 WC-1, WC-2 & FRQ & Neurospora circadian clock\\
 dCLK & PER, TIM & Drospophila circadian clock\\
 CLOCK, BMAL & PER, CRY & Mammals circadian clock\\
 p53 & Mdm2 & Stress response oscillators\\
\hline
\hline
\multicolumn{3}{|c|}{Detected Yeast MFLs}\\
\hline
{Swi6} & {Swi4} & G1/S transition \\
{Gal4} & {Gal80} & Galactose metabolism\\
{Gal4} & {Gal3} & Galactose metabolism\\\
{Gal4} & {Gal1} & Galactose metabolism\\\
{Ime1} & {Rim11} & Meiose activation\\
{Ume6} & {Ime1} & Meiose activation\\
{Ste12} & {Fus3} & Pseudohyphal growth\\
{Ste12} & {Far1} & Pseudohyphal growth\\
{Cbf1} & {Met28} & Sulfur metabolism \\
{Met4} & {Met28} & Sulfur metabolism \\
{Swi4} & {Clb2} & Cell cycle\\
{Mbp1} & {Clb5} & G1/S transition\\
{Stb1} & {Cln1} & \\
{Stb1} & {Cln2} & \\
\hline
\end{tabular}
\caption{Some biological examples of MFL motifs. The yeast motifs are reproduced from ref.~\cite{Yeger04}. An annotation taken from the SGD database
\cite{SGD} has been added.}
\label{bioex.tab}
\end{table}

\begin{appendix}

\section{Transitions between the two phases of an oscillation cycle and
$\sqrt{\delta}$  correction to the period.}
\label{appendixA}

In the following, we analyze the transitions between phase I and II of
an oscillation period and use matched asymptotics to precisely 
justify  the assumptions made in section \ref{osc.sec}. This also provides
the leading order ($\sqrt{\delta}$) correction to the zeroth order results
of section \ref{osc.sec}.

\subsection{From phase I to phase II.}
\begin{figure}
\includegraphics*[width=0.4\textwidth, keepaspectratio=true]{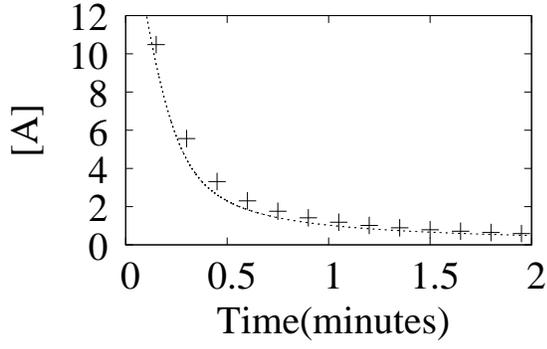}
\caption{\label{match} The evolution of $A$ given by the
Riccati Eq.~(\ref{ric}) (+) is compared to that given by the complete MFL dynamics (dashed)
during the transition from phase I to phase II. Parameters are as in
Fig.~\ref{Oscillation}.
}
\end{figure}

We consider 
the transition between the end of the high A/low B phase I and the start
of the high B/low A phase
II. This protein switch occurs on a time scale of order $\delta/\delta_r$. 
On this fast time scale
the mRNA concentration does not have time to change. Thus, in 
Eqs.~(\ref{gsimple}-\ref{Asimple}), we assume $r=r_2$ and introduce $\tau=t/\delta$.  
Eqs.~(\ref{Bsimple},\ref{Asimple}) simply become at dominant order 
\begin{eqnarray}
\frac{\dd B}{\dd \tau} & {=} & r_2-A\,B\label{Bbl}\\
\frac{\dd A} {\dd \tau} & {=} & 1- A\, B 
\label{Abl}
\end{eqnarray}
The imposed boundary conditions are
\begin{eqnarray}
A&\rightarrow& (1-r_2) (\tau-\tau_1)+o(1),\ B\rightarrow 0\ \mathrm{at}\ \tau=
-\infty\ \ \ \ \ \ \label{bca}\\
B&\rightarrow& (r_2-1) (\tau-\tau_2)+o(1),\ A\rightarrow 0\ \mathrm{at}\  \tau\rightarrow +\infty\ \ \ \ \ \  \label{bcb}
\end{eqnarray}
with the dimensionless transcript concentration $r_2>1$ 
(see section \ref{p1ha}) and $\tau_1$ and $\tau_2$ two constants to be 
determined.
Eqs.~(\ref{Bbl},\ref{Abl}) are autonomous in time and therefore invariant
by time translation. For general non integrable equations,
numerical integration would  be required to obtain the
difference $\tau_2-\tau_1$. Here, however, the difference of protein 
concentrations, $A-B$, is easily integrated across the transition region.
Fixing the time origin at the instant when $A=B$, one readily obtains the exact
formula
\begin{equation}
 A-B=(1-r_2)\tau. 
\label{exf}
\end{equation}
Comparison with Eqs.~(\ref{bca},\ref{bcb}) shows 
that with this choice of time origin 
$\tau_1=\tau_2=0$ (and more generally $\tau_1=\tau_2$). The asymptotic
conditions (\ref{bca},\ref{bcb}) coincide with the limiting behavior of
$a_I(t)/\delta$ at the end of phase I near $t=t_1$, and with the limiting
behavior of $b_{II}(t)$ at the beginning of phase II near $t=0$, provided that
$a_I(t_1)=b_{II}(0)=0$ as was required in the main text. Thus, Eqs.~(\ref{Bbl},
\ref{Abl}) provide a uniform approximation of $A(t)$ and $B(t)$ throughout
the transition from phase I to phase II. Given the exact result (\ref{exf}),
the integration of Eqs.~(\ref{Bbl},
\ref{Abl}) can in fact be replaced to the integration of the single
following Riccati equation:
\begin{equation}
\frac{\dd A}{\dd \tau}=1-[A+(r_2-1)\tau]A
\label{ric}
\end{equation}
A comparison between Eq.~(\ref{ric}) and full numerical evolution is shown in
Fig.~\ref{match}.
There is one subtlety, however. The quasi-equilibrium approximation for $g$ [Eq.~(\ref{gad})] diverges
at the end of phase I when $A \rightarrow 0$ and  $g$ becomes larger
than 1 before the transition region (\ref{Bbl},\ref{Abl}) of order $\delta$. 
This clearly signals the breakdown of the  approximation (\ref{gad}) in a larger
intermediate region at the end of phase I. In a region of size 
$t_1-t\sim \sqrt{\delta}$,
the
evolution of $g$ reduces to
\begin{equation}
\frac{\dd g}{\dd t}  =  \tilde{\theta} \left(1-\,g\frac{a}{\delta A_0}\right)  \label{ginter}
\end{equation}
Indeed,
taking $a=(1-r_2)(t_1-t)+ o(\sqrt{\delta})
$ shows that all three terms kept in Eq.~(\ref{ginter}) are
of the same magnitude when $g\sim t_1-t\sim\sqrt{\delta}$ and that the 
additional term $\tilde{\theta}g$ in Eq.~(\ref{gsimple}) is negligible.
Explicit integration in this intermediate region gives
\begin{equation}
g(t_1)-g(t)\exp\left[-\kappa_2(t_1-t)^2\right]=\tilde{\theta} \int_{t-t_1}^0\!\!
\!\!\dd u
 \exp\left(\kappa_2 u^2\right)
\end{equation}
where $\kappa_2=\tilde{\theta} (1-r_2)/(2 A_0\delta)$.
In particular, the limit $t\rightarrow-\infty$, determines
the value of $g$ at the end of phase I and beginning of phase II

\begin{equation}
g(t_1)=g_2=\sqrt{\frac{\pi A_0\tilde{\theta}\delta}{2(r_2-1)}}
\label{gf1}
\end{equation}

Eq.~(\ref{gf1}) is the source of a correction of order $\sqrt{\delta}$
to the asymptotic result (\ref{periode}) as will be shown below.

\subsection{From phase II to phase I.}
The transition from phase II to phase I can be analyzed quite similarly
to the transition from phase I to phase II. In a time of order
$\delta/\delta_r$ at the end of phase II,  $r$ has no time to change and 
the transition is described by
Eqs.~(\ref{Bbl},\ref{Abl}) with $r_2$ replaced by $r_1<1$ (see section \ref{p1ha}) and the boundary
conditions
\begin{eqnarray}
B&=&(r_1-1) (\tau-\tau_1')+o(1),\ A\rightarrow 0 
\ \mathrm{at}\ \tau= -\infty\ \ \ \ \ \ \ \label{bcb2}\\
A&=&(1-r_1) (\tau-\tau_2')+o(1),\ B\rightarrow 0\ \mathrm{at}\
 \tau= +\infty\ \ \ \ \ \ \ \label{bca2}
\end{eqnarray}
Again, $\tau_1'=\tau_2'=0$ if the time origin is taken when $A=B$. This 
transition regime is followed by a longer transition on a time scale
$t-t_2\sim\sqrt{\delta}$ where $g$ decreases from a value of order one to
values of order $\delta$ characteristic of phase I. This reflects the fact
that the binding of protein A to the promoter of 
gene $g_b$ is not instantaneous and requires some
accumulation of protein A at the beginning of phase I.
Taking $A=(1-r_1) t/\delta$, $g$ evolution is described in this intermediate regime by
\begin{equation}
\frac{\dd g}{\dd t}  =  \tilde{\theta}\left[1-t (1-r_1) \frac{g}{A_0\delta}\right]  \label{ginter2}
\end{equation}
The corresponding evolution of $g$ is
\begin{equation}
g(t)=\exp\left(-\kappa_1 t^2\right)\,
\left[g_1+\tilde{\theta}\int_0^{t}\!\!\!
du\, \exp\left(\kappa_1 u^2\right)\right]
\label{gi2}
\end{equation}
where $\kappa_1=\tilde{\theta} (1-r_1)/(2 A_0\delta)$.
It indeed describes the transition from
 $g_1=1-\exp(\tilde{\theta}t_2)$, the value of $g$ at the end
of phase II, to the quasi-equilibrium regime since 
the asymptotic
behavior of
Eq.~(\ref{gi2}) 
for $t\gg\sqrt{\delta}$ is
\begin{equation}
g(t)\sim \frac{A_0 \delta}{(1-r_1)t}
\end{equation}
\subsection{Dominant correction to the oscillation period.}
The form of Eqs.~(\ref{gsimple}-\ref{Asimple}) could lead to think that
the oscillation period has an expansion in powers of the small
parameter $\delta$. However, here as often, boundary layers lead to more
complicated expansions. The two transition regimes of duration
$\sqrt{\delta}$ give rise to a dominant correction of order $\sqrt{\delta}$
to the oscillation period.

As pointed out above, due to the transition regime at the end of phase I, 
the starting value of $g$ in phase II is $g_2$ [Eq.~(\ref{gf1})] of
order $\sqrt{\delta}$. Therefore, the first Eq.~(\ref{fcompb}) is replaced at order $\sqrt{\delta}$ by the corrected evolution,
\begin{equation}
g_{II,c}(t)=1-(1-g_2)\exp(-\tilde{\theta} t)\label{gm}
\end{equation}
This leads in turn to corrected evolutions $r_{II,c}(t)$ and $b_{II,c}(t)$
for the mRNA and B protein concentrations
that are obtained by simply replacing
$(\rho_1-\rho_0)$ by $(\rho_1-\rho_0)(1-g_2)$ in the expressions 
(\ref{fcompb}) of the zeroth-order expressions $r_{II}(t)$ and $b_{II}(t)$.

The other transition regime between phase II and phase I does not lead to
modifications of the form of Eq.~(\ref{rphase1},\ref{wphase1}) 
describing the evolutions of
mRNA and A protein during the bulk of phase I. Instead, it results in
a difference of order 
$\sqrt{\delta}$ between $r_{II,c}(t_2)$ and $r_1$ the  effective
concentrations
of transcripts at the start of phase I. The transcript concentration depends
on the promoter states at previous times. So, integrating Eq.~(\ref{rsimple})
from the end of phase II, one obtains
\begin{eqnarray}
r(t)&=&\rho_1+[r_{II,c}(t_2) -\rho_1]\exp(-t)
\nonumber\\
&+&(\rho_0-\rho_1)\exp(-t)\int_0^t\!\! \dd u
\exp(u)\,g(u)
\label{rex}
\end{eqnarray}
Replacing $g(t)$ by its expression $(\ref{ginter2})$ during 
the transition regime at the start of phase I leads for $t\gg\sqrt{\delta}$
to
\begin{equation}
r(t)=\rho_1+[r_{II,c}(t_2)-g_1\sqrt{\frac{\pi}{4\kappa_1}}(\rho_1
-\rho_0) -\rho_1]\exp(-t) 
\label{r1c}
\end{equation}
Namely, the previous Eq.~(\ref{rphase1}) but with
\begin{equation}
r_1=r_{II,c}(t_2)-g_1\sqrt{\frac{\pi}{4\kappa_1}}(\rho_1-\rho_0)
\label{r12c}
\end{equation}
Note that only the $g_1$-term between the square brackets in Eq.~(\ref{ginter2}) contribute to Eq.~(\ref{r1c}), the integral term giving a subdominant contribution. 

Finally, the durations $T_1$ and $T_2$ of both phases and the total
period $T_r=T_1+T_2$ are obtained to $\sqrt{\delta}$ accuracy by solving as before
\begin{equation}
r_I(t_1)=r_2,\ a_I(t_1)=0
\end{equation}
together with
the corrected
version of Eq.(\ref{per2.eq}), namely Eq.~(\ref{r12c}) and $
b_{II,c}(t_2)=0$ . 

\begin{figure}
\includegraphics*[width=0.4\textwidth]{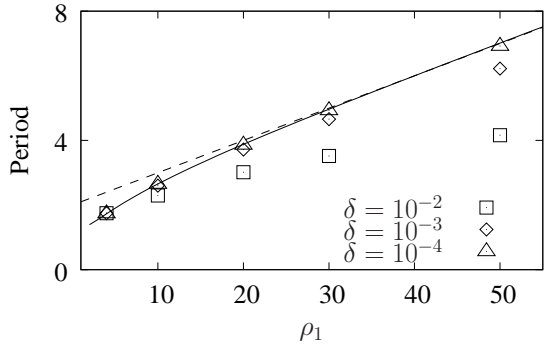}
\caption{\label{period} Comparison between the lowest order theoretical 
rescaled period
$T_r$
(bold line), the approximate expression for large $\rho_1$
[Eq.~(\ref{Periodeapprox})] (dashed line) and
numerically computed
periods for different values of $\delta$, as a function of parameter
$\rho_1$.
Other parameters are $\tilde{\theta}=10$, $A_0=10$, $\mu=1$, $d_a=d_b=0$. }
\end{figure}

\begin{figure}
\includegraphics*[width=0.45\textwidth, keepaspectratio=true]{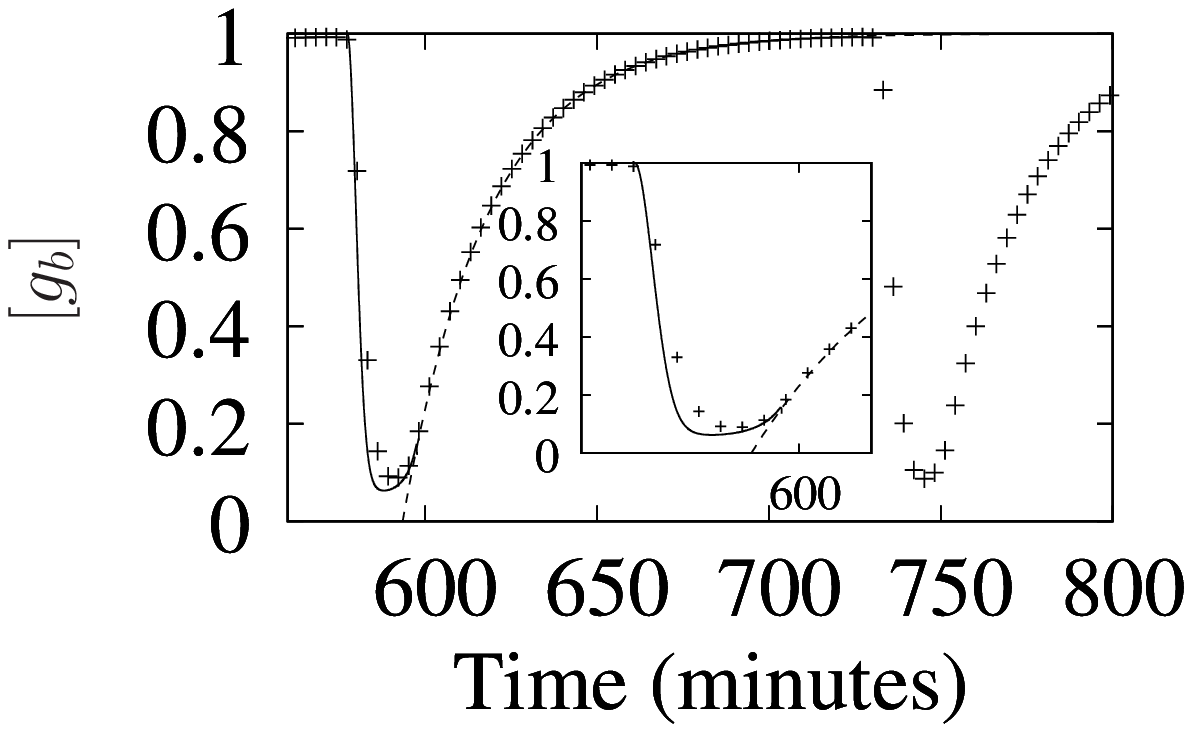}
\caption{\label{gb} Comparison of $[g_b]$ obtained  from a full
numerical integration (+) with the uniform approximation obtained by matching
the different transition regimes (bold line). The exponential relaxation
in phase II [Eq.~(\ref{gm})] is also shown (dashed). An enlargment of phase I and
the transition regimes is shown in the inset.
Parameters are as in
Fig.~\ref{Oscillation}.
}
\end{figure}

The asymptotic description of $g_b$ evolution agrees well with numerical
simulations of the full MFL model as shown in Fig.~\ref{gb}. The $\sqrt{\delta}$ correction terms lengthen the zeroth-order asymptotic estimation. For moderate
value of $\delta$, a numerically comparable contribution is however coming from
higher-order terms that tend to lengthen the period
[coming, for instance, from the breakdown of the 
quasi-equilibrium approximation for $g_b$ Eq.~(\ref{gad})] as can be seen in Fig.~\ref{period}.

\end{appendix}

\newpage

\begin{figure}
\centerline{\begin{tabular}{lr}
\raisebox{100pt}{(a)}&\includegraphics*[width=0.4\textwidth, keepaspectratio=true]{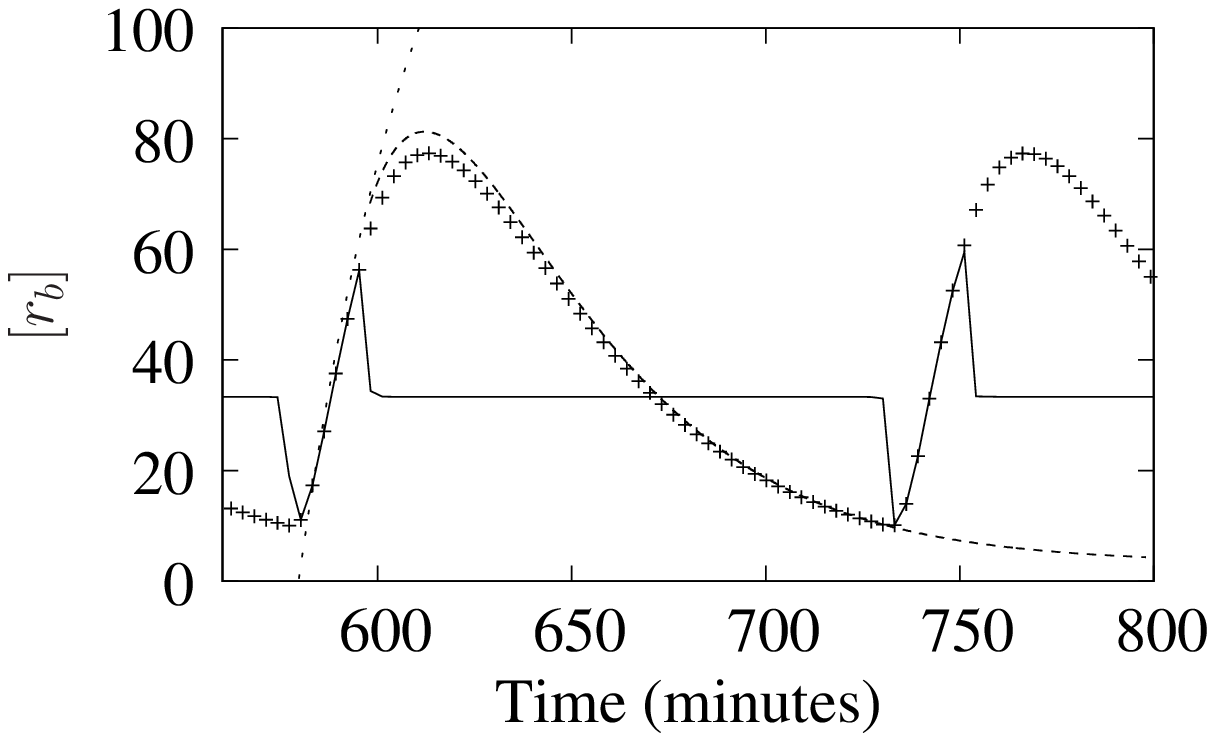}\\
\raisebox{100pt}{(b)}& \includegraphics*[width=0.4\textwidth,
  keepaspectratio=true]{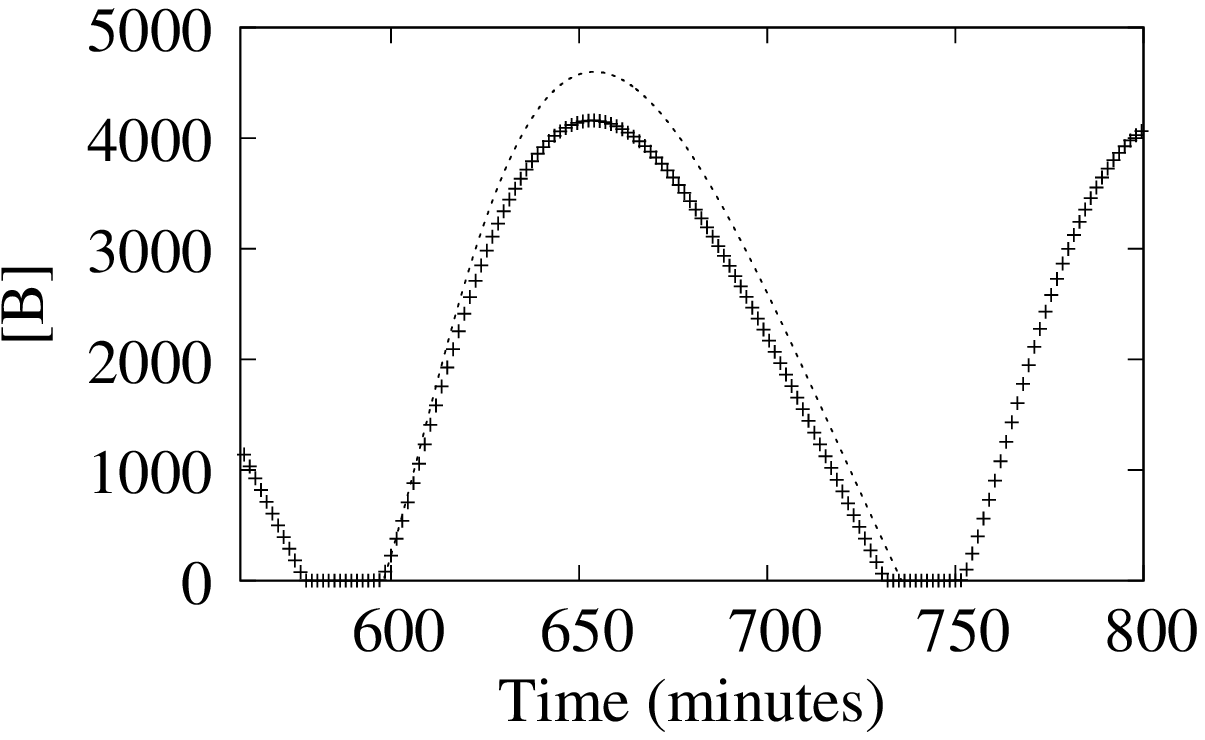}\\
\raisebox{100pt}{(c)}& \includegraphics*[width=0.4\textwidth, keepaspectratio=true]{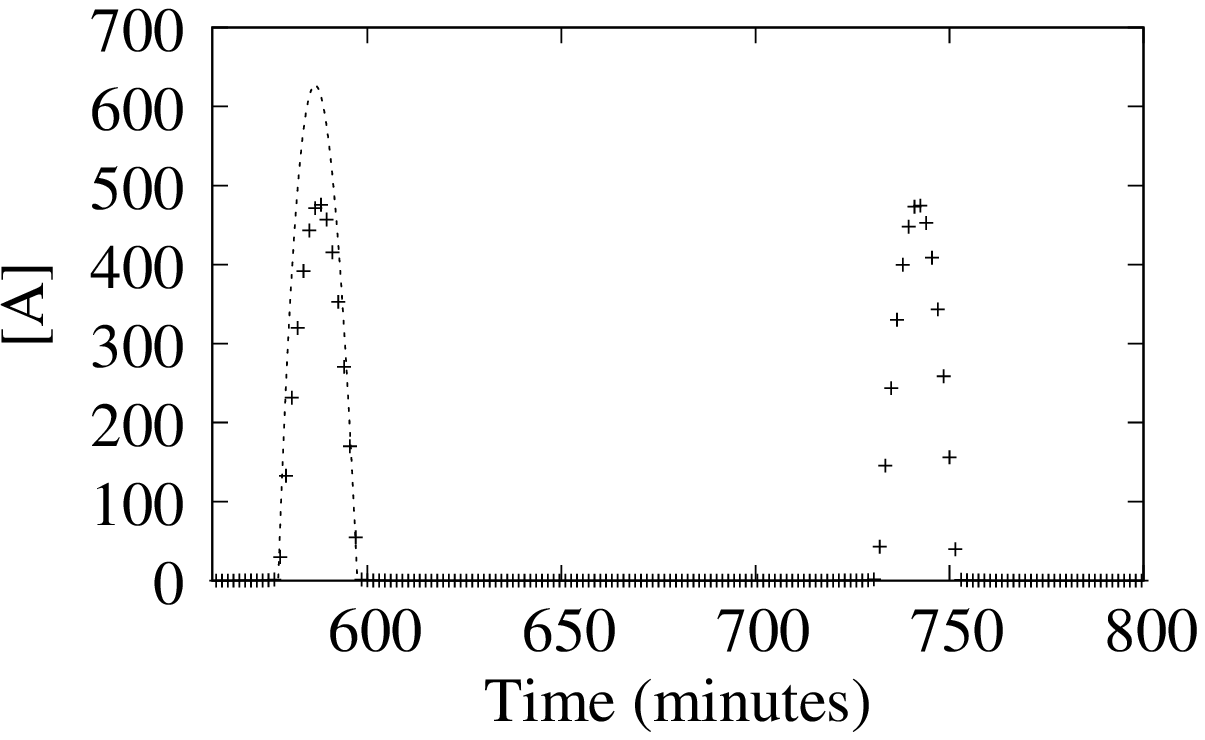}
\end{tabular}}
\caption{
\label{comparison} Comparison between the full dynamics ($+$) symbols
and the asymptotic description at order $\sqrt{\delta}$ (dashed and dotted lines) during one oscillation cycle.
  {\it (a)} Concentration of
 mRNA behavior . Phase I asymptotics (dotted lines) and phase II
asymptotics (dashed lines) are shown separately.
The product
$A.B/\beta$ is also plotted (full line).
 The quasi-static assumption is seen to be
well-satisfied away from the transition regions between the two phases.
{\it (b)} Concentration of protein B.
{\it (c)}  Concentration of protein A.
 Parameters are as in Fig. \ref{Oscillation}.
}
\end{figure}

\begin{thebibliography}{26}
\bibitem{footdis}
These assumptions simplify the analysis but do not play a crucial role.
It was for instance checked that dissociation of the complex with a rate
$\gamma_d$ 
can be accurately described
with the present model
by using an effective  association rate $\gamma$ as briefly
discussed at the end of section \ref{oscsur}.
\bibitem{foot} 
All roots have a
negative real part for the
third degree
polynomial equation with positive coefficients, $X^3+b_1 X^2+b_2 X+b_3=0$,
iff $b_1 b_2-b_3>0$.
\bibitem{Lewfoot} It is ingeniously argued
in 
ref.~\cite{Lewis2003} that $\kappa$ is exactly equal to 2, independently of $n$.
The argument unfortunately appears erroneous. This can be
explicitly checked in the limit $n\rightarrow +\infty$ where the repression
is given by a simple step function and the limit
cycle can be determined as well as $\kappa=\ln[r^2/(r-1)]$ with
$r=\rho/(B_0 c)>1$.
\end{thebibliography}
\end {document}